\documentclass[12pt,twoside]{article}
\usepackage{amsfonts}
\usepackage{amssymb}
\usepackage{amsmath}
\usepackage{appendix}
\usepackage{geometry}
\usepackage{footmisc}
\usepackage{hyperref}
\usepackage{graphicx}
\usepackage{enumerate}
\usepackage[bf]{caption}
\usepackage[space]{grffile}
\usepackage{color}
\usepackage[authoryear,sort]{natbib}
\usepackage[titles]{tocloft}
\usepackage{nc_Format_forSWP55}
\usepackage[french,english]{babel}

\usepackage{pdfpages}

\renewcommand{\footnoterule}{  \kern -3pt  \hrule width 2in  \kern 8pt }
\makeatletter
\renewcommand{\@cite}[1]{#1}
\def\@biblabel#1{\hspace*{-\labelsep}}
\makeatother

\newcommand{\ShortTitleName}{A Semi-Structural Model with Household Debt}
\newcommand{\TitleName}{A Semi-Structural Model with Household Debt for Israel}

\begin{document}
\includepdf[pages=-]{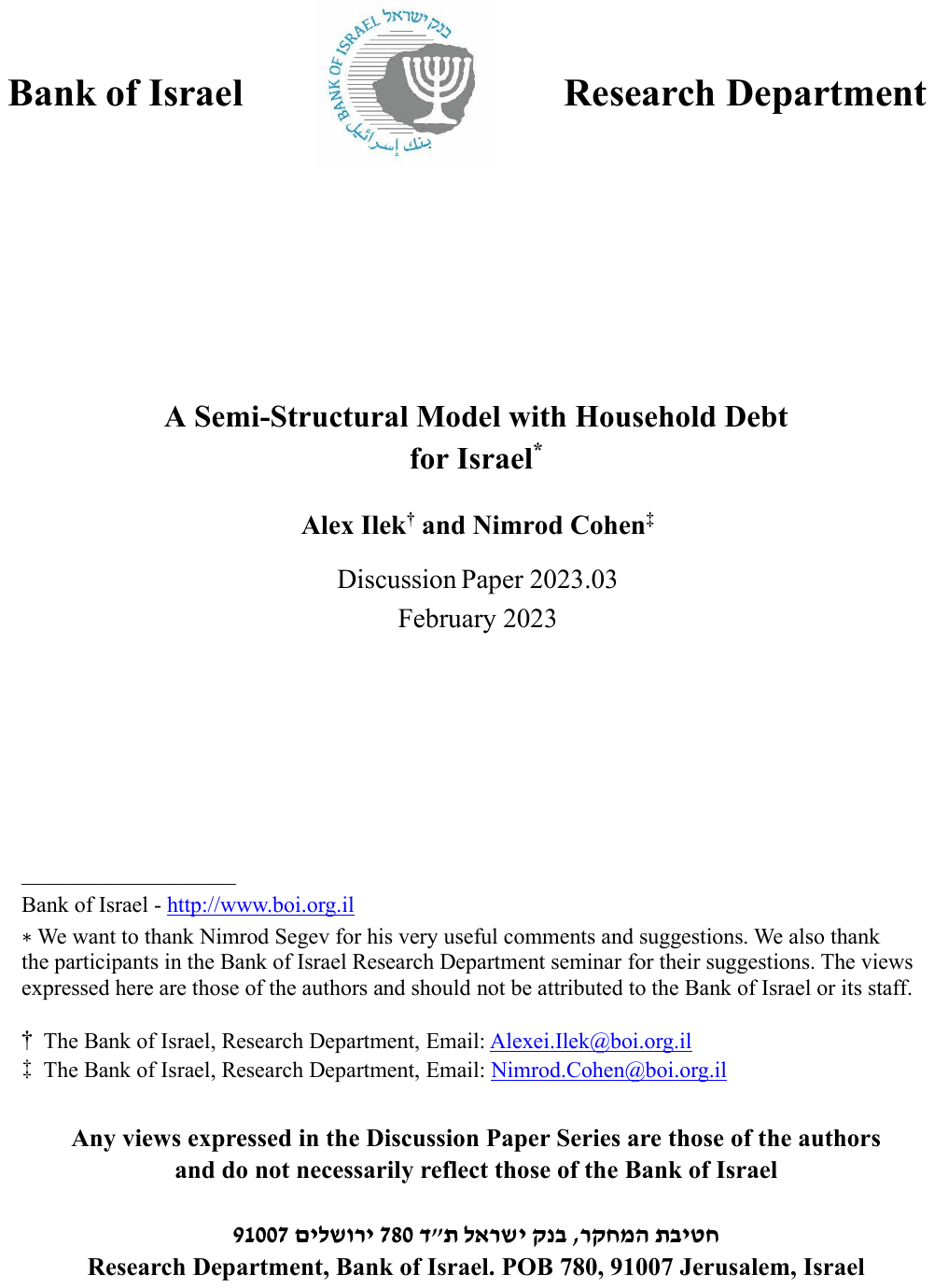}

\title{%
\addcontentsline{toc}{section}{\ShortTitleName}
\textbf{%
\color{\myTitleColor}{\LARGE \TitleName}%
}}
\author{Alex Ilek and Nimrod Cohen 
}
\date{February 7, 2023}
\maketitle

\vspace{1cm}

\begin{abstract}
We propose a semi-structural DSGE model for the Israeli economy, as a small
open economy, which contains a financial friction in the household sector
credit market. Such a friction is reflected in a positive relationship
between households' leverage ratio and their interest rate (credit spread)
on debt, as evident in the Israeli data. Our main purpose is to evaluate the
implications of such a friction on the implementation of monetary policy and
macroprudential policy. Our two main findings are: First, it is important that the monetary policy will react also to developments in the credit market, such as credit spread widening, to increase effectiveness in achieving its main goals of stabilizing inflation and real activity. Second, macroprudential policy may increase the sensitivity
of households' credit spread to their leverage. Thus, this policy can
mitigate or even prevent over-borrowing and reduce the risk of a debt
deleveraging crisis. Moreover, in a case of demand weakness and debt
deleveraging, in addition to accommodative monetary policy, the
macroprudential policy may contribute to stimulating demand due to a
corresponding reduction in credit spread. \newline
\end{abstract}

\setlength{\baselineskip}{22pt}

\runningheads{}{}

\begin{figure}
\centering
\includegraphics[trim=3cm 8cm 3cm 3cm,clip,width=\textwidth]{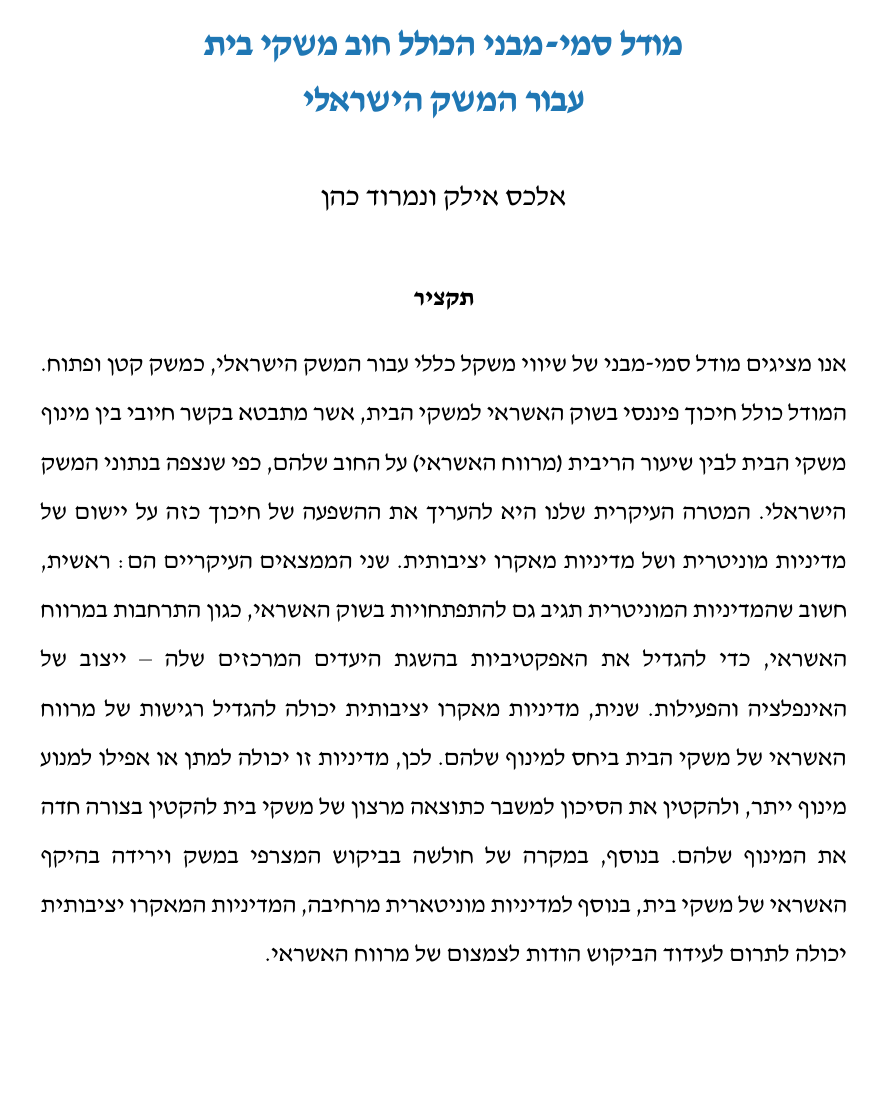}
\end{figure}
\clearpage%

\section{Introduction}

We propose a simple empirical -- semi-structural\footnote{%
A semi-structural general equilibrium model where the structure of the
equations follows economic reasoning, but several economic restrictions are
relaxed to improve the empirical fit and allow for an intuitive use of the
model, as \cite{Laxton_et_al2006}.} -- DSGE model for Israel, as a small
open economy (henceforth, SOE), that incorporates financial friction in the
household sector credit market. We specify and calibrate the model based on
stylized facts and empirical evidence for the Israeli household sector
credit market. Our model incorporates two main blocks. The first block is a
standard model for SOE based on \cite{Laxton_et_al2006}, in which financial
friction is absent, but it contains all the key features of SOE.

The second block contains household sector credit market with financial
friction (for a closed economy). We adopt the type of financial friction
model introduced by \cite{Benigno_et_al_2020} and \cite{Curdia_Woodford_2016}%
, which is reflected in a positive relationship between households' debt
leverage level and their credit interest rate spread. One possible
interpretation for such a relationship is that a higher leverage leads to
higher risk of default perceived by the financial intermediaries, and
therefore requires compensation in the form of spread over a risk-free
interest rate. An alternative interpretation is that providing credit is
costly for the financial intermediaries, and the spread compensates them.

\begin{figure}
\caption{\textbf{Mortgage leverage vs. spread over time in Israel (2004-2022)}}
\includegraphics[width=\linewidth]{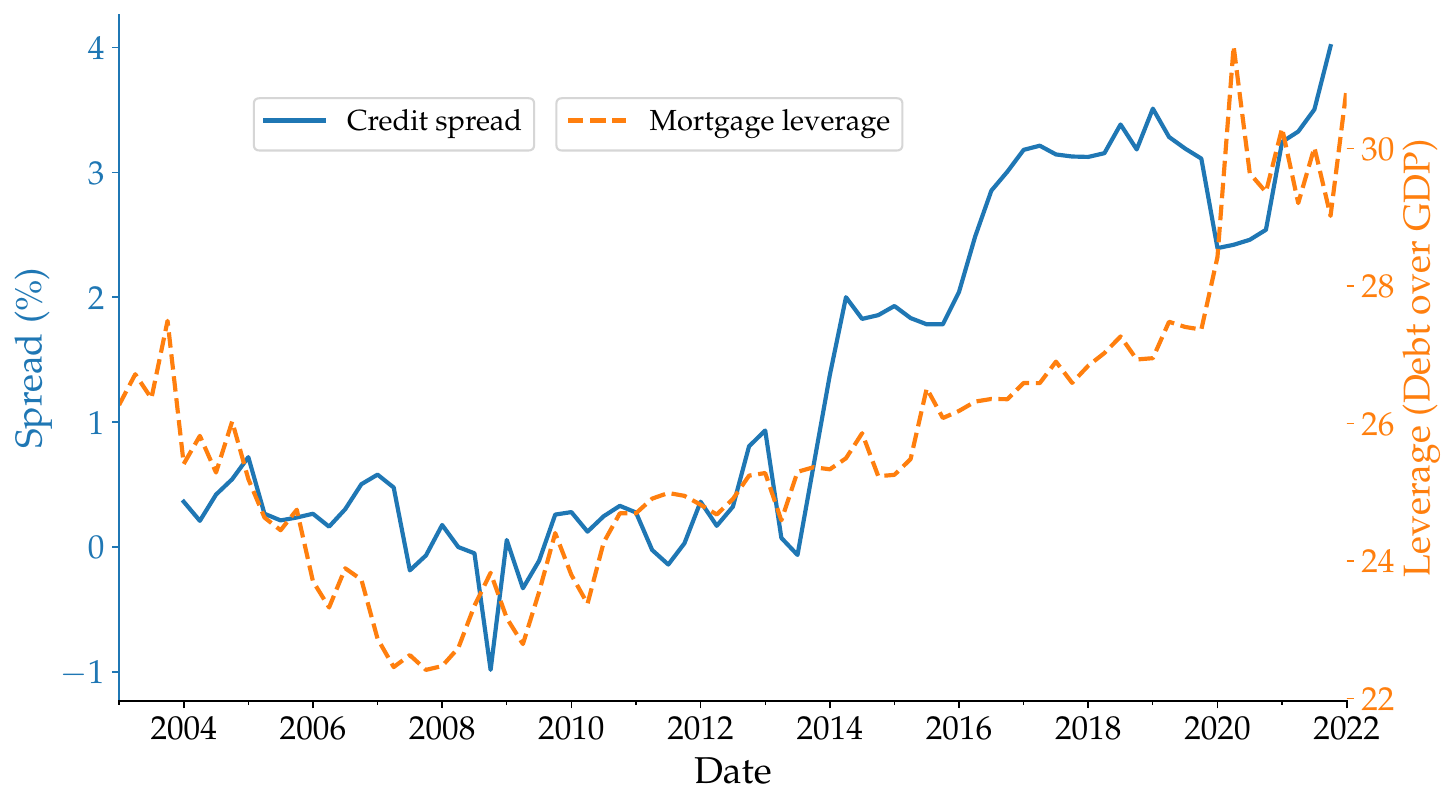}
\label{fig: Spread Leverage}
{\footnotesize
\textit{Notes:} Orange line: leverage defined as mortgage stock over output, 
Blue line: spread defined as the mortgage weighted interest rate over capital market long term bonds (10 years). 
(Source: Bank of Israel)
}
\end{figure}%
\begin{figure}
\caption{\textbf{Average mortgage nominal fixed interest rate over time in Israel (2011-2022)}}
\includegraphics[width=\linewidth]{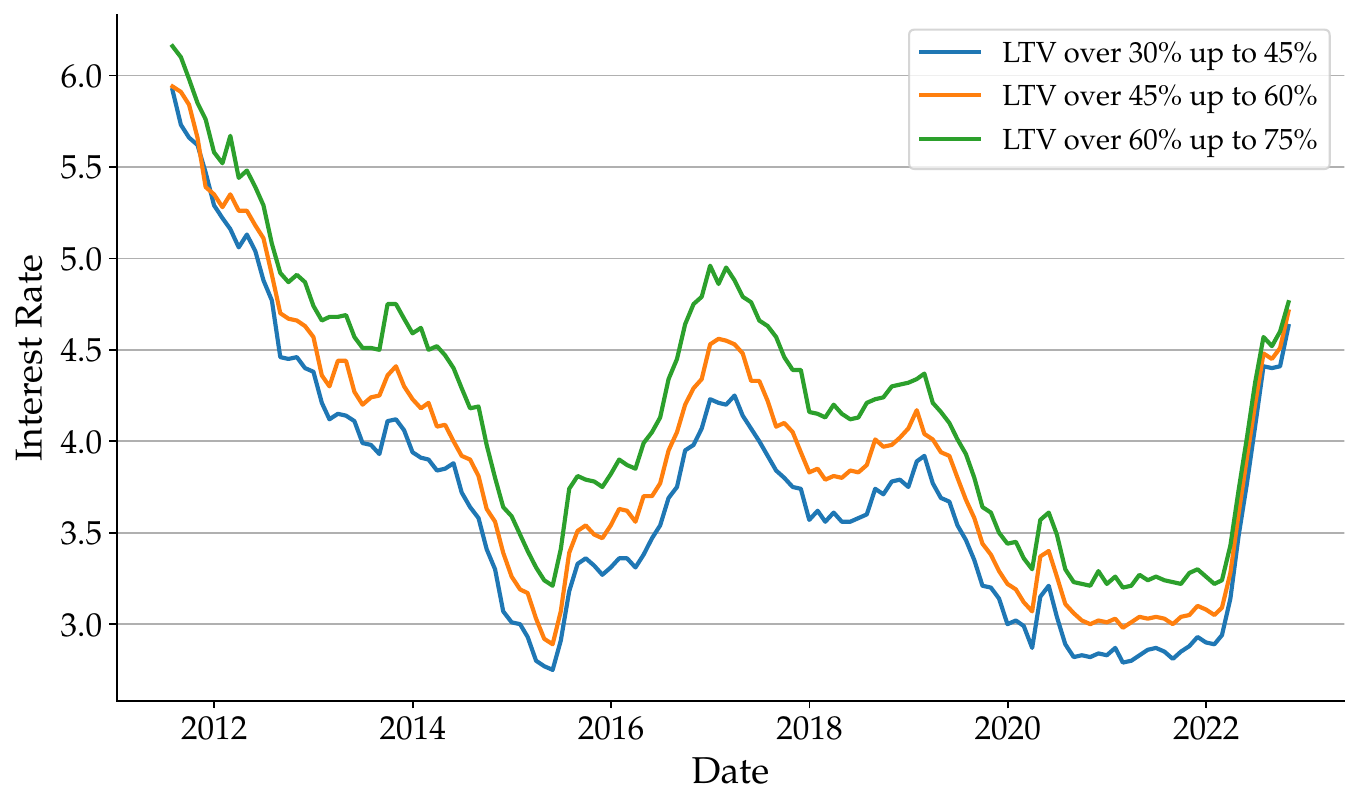}
\label{fig: LTV Spread}
{\footnotesize
\textit{Notes:} Average Mortgage Nominal Fixed Interest Rate, for various LTV groups.
We can see a positive relationship between credit spread and leverage. (Source: Bank of Israel)
}
\end{figure}%

\textbf{Stylized Facts. }We will discuss here about stylized facts and
evidence of financial friction in the household sector credit market in
Israel. First, the ratio of borrowers in Israel is about 40\% (according to 
\cite{Shami_2019}), which motivates a model with heterogeneous households --
two types of agents -- borrowers and lenders. Second, the Israeli household
sector credit market is closed to abroad, and just a minute percent of the
households can borrow from abroad. Moreover, their dominant credit is
mortgage, which is around two-thirds of their total debt, on average, over
time. Finally, to justify integration of the type of financial friction we
describe above into our model, we empirically validate the existence of a
positive relationship between the interest rate spread and leverage ratio
(as we will define below) in the Israeli mortgage market. Note that the
Israeli data on household's non-housing debt interest rates are too short
(started only in 2016)\ such that we cannot draw a conclusion about their
credit spread--leverage relation. Fig. \ref{fig: Spread Leverage} presents
the leverage in the Israeli mortgage market defined as mortgage stock over
output (orange line), and the corresponding credit spread defined as the
mortgage weighted interest rate over capital market long term bonds (10
years), in the blue line. We can clearly see a positive relationship between
the two -- mortgage leverage and credit spread -- over time (2004-2022).
Another indication for a positive relationship can be observed in Fig. \ref%
{fig: LTV Spread}, which presents the average mortgage nominal fixed
interest rate over time, for various LTV groups. As expected, we can see
that as LTV is higher the spread is higher. In summary, these enlightening%
\textsf{\ }figures support the existence of financial frictions in Israel's
credit market. Formal analyses can be seen in Section \ref{subsubsec: spread
elasticity} and Appendix \ref{app: empirical evidence}.

We will now elaborate on the details of the financial friction modeling
framework. In \cite{Benigno_et_al_2020} and \cite{Curdia_Woodford_2016}, the
leverage ratio is defined by current aggregate households' debt over the 
\textit{steady state output}. We will adjust this definition to be
households' debt over their income (which in turn is related to the current 
\textit{output gap}), as we think that this definition is more suitable in
practice.\footnote{%
Practically, it is common that when a borrower applies to receive a
mortgage, the bank requires an income report at least to evaluate the
ability of the borrower to repay monthly payments and minimize the
probability of default.} Moreover, with this definition, our model can
produce either a financial accelerator or financial decelerator dynamic,
depends on the parameters' calibration. We will elaborate on this in Section %
\ref{subsubsection: Financial Accelerator or Financial Ddecelerator} (and
Appendix \ref{app: IRF Financial Accelerator}), and we will validate that
the current model may fit well to the Israeli case in Section \ref%
{subsubsec: the case of Israel}.

In fact, the definition of leverage we use is closely related to the
definition in \cite{Bernanke_et_al_1999}, \cite{Gertler_Karadi_2011} and 
\cite{Iacoviello_2005} in the sense that the 
\foreignlanguage{french}%
num\'{e}raire 
\foreignlanguage{english}%
of the leverage takes into account the value of economic unit or value of
asset. In \cite{Bernanke_et_al_1999} and \cite{Gertler_Karadi_2011} the 
\foreignlanguage{french}%
num\'{e}raire 
\foreignlanguage{english}%
is a firm's (bank's) net worth (value of assets). At \cite{Iacoviello_2005}
the 
\foreignlanguage{french}%
num\'{e}raire 
\foreignlanguage{english}%
is the value of house, which is a collateral for the lenders. In our
framework, the 
\foreignlanguage{french}%
num\'{e}raire 
\foreignlanguage{english}%
is the household's income that captures to some extent the present (or even
expected) income of the borrowers. An increase in the household's income,
ceteris paribus, reduces risk of default and should reduce the cost of
borrowing. Thus, increase in debt does not necessarily induce higher spread,
unless income increases by a lower extent.

One of the most striking results in \cite{Benigno_et_al_2020}, which we
inherit in our model, is that the natural rate of interest (henceforth, NRI)
is negatively affected by households' debt (or by interest rate spread).
Thus, the NRI is partially affected by policy because the households' debt,
as well as interest rate spread, are policy dependent. Since the NRI is
included in the aggregate demand equation (IS), a higher spread leads to
lower aggregate activity. Thus, the model links negatively financial
conditions and real activity. A negative effect of financial conditions on
real activity was also presented (for a closed economy) by \cite%
{Adrian_et_al_2020}\footnote{%
In \cite{Adrian_et_al_2020} which is not microfounded, the negative
relationship is between a financial condition index and output gap, but the
main economic essence is the same.}, and by \cite{Curdia_Woodford_2016}%
\footnote{%
Although in their model the negative effect stems from the expected flow of
future spreads rather than from a one-period spread.\ To be precise, the
expected future flow of spreads at period $t$ negatively affects the output,
but the expected future flow of spreads at period $t+1$ positively affects
output gap.}.

It should be noted that \cite{Benigno_et_al_2020} obtained this relationship
from microfoundations, therefore the NRI in their model is a real interest
rate that would have been obtained under price flexibility. Since our model
is not fully microfounded, we cannot necessarily treat\ the NRI here
according to the same definition as in \cite{Benigno_et_al_2020}.
Nevertheless, we can still treat it as \textbf{a benchmark real interest rate%
} for the central bank (henceforth, CB). Only for the sake of simplicity,
from now on we will call the benchmark interest rate NRI.

Furthermore, the NRI in our model for SOE is also affected by the expected
growth of potential output in the domestic economy, the expected growth
abroad (see \cite{CGG_2002}), and the demand of borrowers for credit
(because of shocks to their preference).

Given our model, we will ask three main questions: (1) What are the
implications of the financial friction -- in the household sector credit
market -- on the Israeli economy under various types of shocks? (2) Should
monetary policy react to financial variables, like credit spread or
leverage? What are the costs, in terms of efficiency of monetary policy to
achieve its main goals -- price and real activity stability, if the monetary
policy ignores the financial sector? (3) What are the implications of
macroprudential policy that seeks to mitigate excess lending?

These questions are closely related to a still open debate in the economic
literature -- the \textit{Leaning Against the Wind} debate -- which
intensified after the global financial crisis in 2008, asking whether
monetary policy should respond to financial variables or not (\cite%
{Boissay.etal2022Monetary} and \cite{Gourio.etal2018Trade}). On one side,
economists suggest the need for tighter monetary policy for financial
stability purposes -- Leaning Against the Wind policy (henceforth, LAW) --
see \cite{Borio2014Macroprudential}\footnote{%
and \cite{Juselius.etal2016Monetary}, \cite{Borio.etal2018Monetarya}.}, \cite%
{Curdia_Woodford_2016}, \cite{Curdia_Woodford_2010} and \cite%
{Adrian_et_al_2020}. Based on the idea that high level of borrowing
significantly increases the probability of future default and may have a
potential disastrous effect on the economy, this calls for policy
intervention to control the financial risk. Monetary policy may mitigate to
some extent the excess borrowing by increasing the interest rate. However,
attaining financial stability with monetary policy may contradict achieving
stability of inflation and real activity. This is why on the other side,
some economists claims that each policy -- monetary policy and
macroprudential policy -- should be conducted separately, with its separate
goals and instruments (the \textit{separation principle} of \cite%
{Svensson2018Monetary}\footnote{%
and \cite{Svensson2014Inflation}, \cite{Svensson2017Costbenefit}.}). Thus,
only macroprudential policy can deal directly with financial stability
issues.

The rest of the paper is organized as follows. Section \ref{Section: Model}
presents the main equations of the model, Section \ref{Section: Calibration}
describes the data used for validating the existence of the financial
friction in Israel, and presents the calibration of the model's parameters,
which is most relevant for the financial sector, Section \ref{Section:
Implication of FF} shows the implication of the financial friction on the
economy, and also shows the implications of macroprudential policy. Section %
\ref{Section: Monetary policy analysis} presents the analysis of monetary
policy, and Section \ref{Section: Conclusions} concludes the main results.

\section{The Model\label{Section: Model}}

Our model consists of domestic and foreign parts. The foreign part
introduces the world economy model which has a New-Keynesian standard
specification for a closed economy. The domestic part is presented by the
SOE economy with a financial sector for households' credit, as discussed
above. Moreover, in the current model, the SOE has two types of households
-- borrowers and lenders -- in contrast to \cite{Laxton_et_al2006} where the
SOE model is a representative agent model without a financial sector. Note
that in our accomplished model we do not explicitly include lenders
equations, since the model explicitly presents the aggregate activity and
the borrowers' activity, therefore lenders' activity is redundant.

The main model advantage is that although it is parsimonious and empirically
oriented (for the Israeli economy), it is capable of answering basic
economic questions regarding the implications of financial friction on
aggregate activity and other macroeconomic variables in Israel. However,
since the model is not microfounded, it is limited in its ability to provide
exact answers to all kinds of questions, especially related to implications
of changing deep parameters of the model.

Since our model is for SOE, attention should be given to foreign forces
affecting the domestic economy. First, as we saw above, domestic households
in Israel cannot\textit{\ }borrow from abroad, so their credit is entirely
domestic, therefore the domestic spread and leverage are \textbf{only
indirectly affected} by foreign economy (and foreign shocks). For instance,
in Appendix \ref{app: Foreign shock IRF}, we present the responses of the
main macroeconomics variables - output gap and inflation - to foreign
monetary shocks, and it is clear that the responses are not sensitive for
change in the domestic households' financial friction. Second, the
possibility for households to invest (and save) abroad has existed in Israel
for many decades, mostly through financial intermediaries. But this is
costly, so the domestic and foreign assets are not complete substitutes, and
thus savers should be indifferent between receiving a higher interest rate
abroad subject to transaction costs or a lower interest rate in the domestic
economy. This incomplete substitutability between domestic and foreign
assets gives rise in our model to a non-zero effect of domestic spread on
the NRI (see Sec.\ref{Section: Calibration}), and also to the existence of
the risk premia in the UIP equation (as in \cite{Schmitt_Grohe_and_Uribe_2003}).

\subsection{The SOE model equations}

The main equations of the non financial block are similar to \cite%
{Laxton_et_al2006}, although we modified some of them. The financial block
equations are based on microfounded equations from \cite{Benigno_et_al_2020}
and \cite{Cohen2022Analyzing}. Subscripts $t+1$ and $t+4$ in the equations
below denote expectation for the next quarter and over the next four
quarters, respectively. We consider three foreign shocks: monetary shock,
inflation shock and potential output growth shock abroad. We also consider
six domestic shocks: demand (preference of the borrowers) and supply
financial shocks, monetary shock, inflation shock, potential output growth
shock, and exchange rate shock.

Below we provide a brief explanation of the model equations.

\begin{enumerate}
\item Phillips curve. The inflation rate is determined by past and expected
inflation (for last year and next year, respectively, as $\pi _{t}^{4q}=%
\frac{1}{4}(\pi _{t}+\pi _{t-1}+\pi _{t-2}+\pi _{t-3})$ is four quarters
inflation), output gap, real depreciation of the shekel, and oil prices. $%
\varepsilon _{t}^{\pi }$ is a cost-push shock (inflation shock).%
\begin{eqnarray}
\pi _{t} &=&A_{ld}^{\pi }\pi _{t+4}^{4q}+(1-A_{ld}^{\pi })\pi
_{t-1}^{4q}+A^{y}\widehat{y}_{t-1}  \label{philips} \\
&&+A_{z}^{\pi }\Delta z_{t}+A_{oil}^{\pi }\pi _{t}^{oil}+A_{oil,lag}^{\pi
}\pi _{t-1}^{oil}+\varepsilon _{t}^{\pi }  \notag
\end{eqnarray}

\item Uncovered Interest Rate Parity equation (henceforth, UIP) for real
exchange rate of the shekel (henceforth, RER). It is determined by expected
and past RER, interest rate spread between domestic and foreign riskless
real interest rates, and country risk premia.%
\begin{equation}
z_{t}=D_{zld}^{z}z_{t+1}+(1-D_{zld}^{z})z_{t-1}-(r_{t}-r_{t}^{\ast }-\Delta
_{t}^{fx})  \label{UIP}
\end{equation}

\item The country risk premia in the UIP is determined by the change in the
natural RER, $z_{t}^{n}$, and the gap between the domestic and world NRI. $%
\varepsilon _{t}^{fx}$ is a shock to country risk premia (see, for example 
\cite{Schmitt_Grohe_and_Uribe_2003}).%
\begin{equation}
\Delta _{t}^{fx}=\Delta z_{t}^{n}+(r_{t}^{n}-r_{t}^{nw})+\varepsilon
_{t}^{fx}  \label{UIP_PREMIA}
\end{equation}

\item Aggregate IS equation. The output gap is determined by expected and
past output gap, monetary stance, $\widehat{r}_{t}$, which is a gap between
the real interest rate and the natural real interest rate ($\widehat{r}%
_{t}=r_{t}-r_{t}^{n}$, where $r_{t}=i_{t}-E_{t}\pi _{t+1}$ is Fisher
equation), real exchange rate gap ($\widehat{z}_{t}=$ $z_{t}-z_{t}^{n}$),
and output gap abroad.%
\begin{equation}
\hat{y}_{t}=\beta _{yld}^{y}\widehat{y}_{t+1}+(1-\beta _{yld}^{y})\widehat{y}%
_{t-1}-\beta _{r}^{y}\widehat{r}_{t-1}+\beta _{z}^{y}\widehat{z}_{t-1}+\beta
_{y^{w}}^{y}\widehat{y}_{t}^{w}  \label{ygap}
\end{equation}%
\textsf{\ \ \ \ }\newline
This equation represents the dynamic of the aggregate output gap that has
the following composition. First, note that the model is without investment
and capital dynamics. Second, \textbf{net export,} in line with \cite%
{Laxton_et_al2006}, can be split into quantity dynamic represented by $%
\widehat{y}_{t}^{w}$ and relative prices dynamic represented by $\widehat{z}%
_{t-1}$. Finally, exluding the above (two right elements in Eq. \ref{ygap})
we remain with the Euler equation for the aggregate consumption (the joint
consumption of two type of agents -- borrowers and lenders -- in line with 
\cite{Benigno_et_al_2020}). This explanation is important for the
implementation of the next equation (NRI).

\item The NRI. Our specification of the NRI is completely different from 
\cite{Laxton_et_al2006} and \cite{Chen_Zion_2021}, who modeled the NRI and
the growth rate of potential output as two independent stationary processes.
Here the NRI consists of four components, as follows. The first two
components are the expected growth of potential output of the SOE and the
expected growth abroad (actual). This specification was introduced by \cite%
{CGG_2002} for an open economy without financial friction. The last two
components capture the effect of the financial sector on the aggregate
activity (through the NRI). Specifically, the third component is the
domestic interest rate spread reflecting financial friction, as was
introduced by \cite{Benigno_et_al_2020}. The minus sign reflects
contractionary effect, since higher spread leads to lower aggregate
activity, in line with the literature (\cite{Adrian_et_al_2020} and \cite%
{Curdia_Woodford_2016}). Moreover,\ \cite{Benigno_et_al_2020} considered
only supply shocks to credit stemming from financial intermediaries, so in
the fourth component we also consider demand shocks for credit, caused by
preference shock of the borrowers\footnote{%
It can be shown that adding preference shocks for the borrowers in the model
of \cite{Benigno_et_al_2020} results in an additional component in the NRI,
which is the expected change in the preference shocks of the borrowers
multiplied by their relative share in the population.}.%
\begin{equation}
r_{t}^{n}=\alpha +\alpha _{g}^{NRI}g_{t+1}^{n}+\alpha
_{g^{w}}^{NRI}g_{t+1}^{w}-\alpha _{\Delta }^{NRI}\widehat{\Delta }%
_{t}-\alpha _{cb}^{NRI}\Delta \varepsilon _{t+1}^{cb}  \label{NRI}
\end{equation}

\item Central bank policy rule. The monetary interest rate is determined by
a Taylor-type rule with interest rate inertia. It reacts to the NRI,
inflation environment, which is a weighted average of deviation from the
target of the expected inflation over the next year and actual inflation in
the past year, and output gap. $\varepsilon _{t}^{i}$ is a monetary shock. 
\begin{eqnarray}
i_{t} &=&G_{lag}^{i}i_{t-1}+  \label{policy_rule} \\
&&(1-G_{lag}^{i})(r_{t}^{n}+\overline{\pi }+G_{\pi }^{i}\left( 
\begin{array}{c}
w(\pi _{t+4}^{4q}-\overline{\pi })+ \\ 
(1-w)(\pi _{t}^{4q}-\overline{\pi })%
\end{array}%
\right) +G_{y}^{i}\widehat{y}_{t})+\varepsilon _{t}^{i}  \notag
\end{eqnarray}

\item The growth rate of the potential output is determined by past growth
of potential output, depreciation of the shekel in terms of the natural RER,
and growth of actual output abroad. Inclusion of the two variables into Eq. %
\ref{pot_output} is also an extension of \cite{Laxton_et_al2006}, and
reflects the world demand for domestic output under price flexibility. The
potential output growth is also affected by shocks ($\varepsilon _{t}^{g}$),
such as technology shocks or fiscal shocks\footnote{%
Since the presented model is not microfounded, it is impossible to
differentiate between all shocks that affect the potential output growth.}.
In steady state (henceforth, SS), the growth of potential output is
positive, $g^{n}>0,$ reflecting long-run technological growth of the economy.%
\begin{equation}
g_{t}^{n}=(1-\theta _{1}-\theta _{3})g^{n}+\theta _{1}g_{t-1}^{n}+\theta
_{2}\Delta z_{t}^{n}+\theta _{3}g_{t}^{w}+\varepsilon _{t}^{g}
\label{pot_output}
\end{equation}
\end{enumerate}

The equations listed below describe the financial sector of the model.

\begin{enumerate}
\item Leverage ratio gap, $\widehat{lev}_{t}$. First, let us define $b_{t}$
and $y_{t}$ as levels of household debt and output, respectively. Define $%
lev_{t}=\frac{b_{t}}{y_{t}}$ as a debt-output leverage ratio. We assume that
in SS, output and debt grow at their potential rates, $g^{n}$, therefore the
leverage ratio is constant and equal to $lev=\frac{b}{y}$. Any deviation of
debt and/or output from their "natural" level leads to a deviation of the
leverage ratio from its SS value. Therefore, we present the leverage ratio
deviation from SS as a difference between debt gap and output gap. 
\begin{equation}
\widehat{lev}_{t}=\widehat{b}_{t}-\widehat{y}_{t}  \label{leverage}
\end{equation}

\item Credit spread (Identity) on the households' debt is the interest rate
spread between the interest rate for borrowers and lenders%
\begin{equation}
\Delta _{t}=i_{t}^{b}-i_{t}  \label{spread}
\end{equation}

\item Credit spread (supply side). Assume that the credit spread is a
function of the leverage ratio, $\Delta _{t}=\beta _{lev}^{\Delta
}lev_{t}+\varepsilon _{t}^{\Delta },$ where $\beta _{lev}^{\Delta }>0.$ $%
\varepsilon _{t}^{\Delta }$ is a financial shock which either presents
changes in the perceived level of "safe" debt by financial intermediates or
changes in costs associated with credit provision by financial
intermediates. In SS, the spread is positive, $\Delta >0$. For further
purposes it is convenient to represent the relationship in terms of gaps,
namely deviations from SS values. 
\begin{equation}
\widehat{\Delta }_{t}=\beta _{lev}^{\Delta }\widehat{lev}_{t}+\varepsilon
_{t}^{\Delta }  \label{spread leverage}
\end{equation}%
One can decompose the total elasticity of the spread with respect to
leverage as $\beta _{lev}^{\Delta }=\beta _{lev}^{inter}+\beta _{lev}^{MP}$,
where $\beta _{lev}^{inter}$ captures only private banks' considerations of
penalizing higher debt, taking into account their own revenues from the
interest rate and risk of default. If for the policymakers $\beta
_{lev}^{inter}$ of banks is perceived to be too low, that is, the financial
risks are undervalued, the intervention can be in place, by increasing the
total elasticity by $\beta _{lev}^{MP}$. This can be done by intervention
that makes the credit more expensive, causes an increase in the elasticity
of the spread to leverage. For example, by increasing the capital
requirements on loans provided by banks, as was implemented in Israel in
2010, where capital requirements for risky loans were set (see detailed
description in \cite{Benchimol_et_al_2021}).

\item \textbf{Borrowers }Euler equation. Their consumption is determined by
expected and past consumption, real interest rate for borrowers, leverage
ratio, and expected growth of potential output. All variables here are
expressed in gaps (i.e. $\widehat{g}_{t}^{n}=g_{t}^{n}-g^{n}$). $\varepsilon
_{t}^{cb}$ is a consumption preference shock of the borrowers which
represents demand shock to credit. 
\begin{eqnarray}
\widehat{c}_{t}^{b} &=&\beta _{ld}^{c^{b}}\widehat{c}_{t+1}^{b}+(1-\beta
_{ld}^{c^{b}})\widehat{c}_{t-1}^{b}-\beta _{r}^{c^{b}}(\hat{R}%
_{t}^{b}-\left( \pi _{t+1}-\overline{\pi }\right) +v~\widehat{lev}_{t}-%
\widehat{g}_{t+1}^{n})  \notag \\
&&-\Delta \varepsilon _{t+1}^{cb}  \label{Euler_borrowers}
\end{eqnarray}

\item \textbf{Borrowers }budget constraint (in real terms). The left hand
side of the equation is the amount of credit at period $t$ needed to finance
several components on the right hand side: (1) repayment of old credit from
period $t-1$ and the interest rate payment on debt in period $t$. Since the
debt is nominal, high inflation in the past erodes the old debt in real
terms reducing required $\hat{b}_{t}$; the growth rate ($\widehat{g}_{t}^{n}$%
) is a by-product of stationarization\footnote{%
The original borrowers budget constraint is $\frac{b_{t}}{R_{t}^{b}}%
=c_{t}^{b}+\frac{b_{t-1}}{\Pi _{t}}-\frac{\varpi }{\chi }y_{t}$. Dividing by 
$y_{t}$ yields $\frac{b_{t}}{R_{t}^{b}y_{t}}=\frac{c_{t}^{b}}{y_{t}}+\frac{%
b_{t-1}}{\Pi _{t}y_{t}}\frac{y_{t-1}}{y_{t-1}}-\frac{\varpi }{\chi }$, where
now the third term is divided by the growth rate $\frac{b_{t}^{y}}{R_{t}^{b}}%
=\frac{c_{t}^{b}}{y_{t}}+\frac{b_{t-1}^{y}}{\Pi _{t}\Delta y_{t}}-\frac{%
\varpi }{\chi }$.}\textit{;}(2) a gap between consumption and income gaps
(we assume that $\beta _{b}^{c^{b}}=\beta _{b}^{y})$.%
\begin{equation}
\widehat{b}_{t}=\widehat{R}_{t}^{b}+\beta _{lag}^{b}\left( \widehat{b}%
_{t-1}-\left( \pi _{t}-\overline{\pi }\right) -\widehat{g}_{t}^{n}\right)
+\beta _{b}^{c^{b}}\widehat{c}_{t}^{b}-\beta _{b}^{y}\widehat{y}_{t}
\label{budget_b}
\end{equation}%
This is consistent with the stylized fact discussed before -- The source of
credit for households is only domestic and it comes from the domestic
lenders (through financial intermediates).
\end{enumerate}

\subsection{The world economy equations}

Most equations of the world economy are from \cite{Laxton_et_al2006}. The
only modification we made here is concerning the specification on the NRI in
Eq. \ref{NRI_w} according to \cite{CGG_2002}.

\begin{enumerate}
\item Inflation equation (Phillips curve)%
\begin{equation}
\pi _{t}^{w}=\alpha ^{w}E_{t}\pi _{t+4}^{4qw}+(1-\alpha ^{w})\pi
_{t-1}^{4qw}+k^{w}\hat{y}_{t-1}^{w}+\varepsilon _{t}^{\pi ^{w}}
\label{philips_w}
\end{equation}

\item IS equation%
\begin{equation}
\hat{y}_{t}^{w}=\delta _{1}^{w}E_{t}\hat{y}_{t+1}^{w}+\delta _{2}^{w}\hat{y}%
_{t-1}^{w}-\delta _{2}^{w}\widehat{r}_{t-1}^{w}  \label{ygap_w}
\end{equation}

where monetary stance abroad is $\widehat{r}_{t}^{w}=r_{t}^{w}-r_{t}^{wn}$,
and Fisher equation $r_{t}^{w}=i_{t}^{w}-E_{t}\pi _{t+1}^{w}$.

\item Natural rate of interest%
\begin{equation}
r_{t}^{nw}=c_{0}^{w}+c_{1}^{w}g_{t+1}^{nw}  \label{NRI_w}
\end{equation}

\item Policy rule of the CB%
\begin{equation}
i_{t}^{w}=\gamma ^{w}i_{t-1}^{w}+\left( 1-\gamma ^{w}\right) \left(
r_{t}^{nw}+\overline{\pi }^{w}+\beta _{1}^{w}(\pi _{t+4}^{4qw}-\overline{\pi 
}^{w})+\beta _{2}^{w}\hat{y}_{t}^{w}\right) +\varepsilon _{t}^{rw}
\label{policy_rule_w}
\end{equation}

\item Growth rate of the potential output%
\begin{equation}
g_{t}^{nw}=(1-\theta ^{w})g^{nw}+\theta ^{w}g_{t-1}^{nw}+\varepsilon
_{t}^{gw}  \label{pot_output_w}
\end{equation}
\end{enumerate}

\section{Calibration\label{Section: Calibration}}

\subsection{Data}

The data regarding credit volume is based on the commercial banks' financial
reports and Bank of Israel process. The credit spread is based on a monthly
report of the commercial bank to the Banking Supervision Department at the
Bank of Israel.

\subsection{Calibration stages}

We summarize the main model parameters in Table 
\ref{table:calibration}%
. Most of the parameters in the non-financial block are based on \cite%
{Chen_Zion_2021} who estimated the model of \cite{Laxton_et_al2006} for
Israel. As we showed in Section \ref{Section: Model}, the effect of the
financial friction on aggregate activity is through the NRI in Eq. \ref{ygap}%
. Therefore, in order to correctly assess the implications of the financial
friction on aggregate activity, three parameters require special attention:
(1) the elasticity of the interest rate spread to debt-GDP ratio, $\beta
_{lev}^{\Delta }$, in Eq. \ref{spread leverage}; (2) the elasticity of NRI
to the spread (up to minus sign), $\alpha _{\Delta }^{NRI}$, in Eq. \ref{NRI}%
; and (3) the elasticity of the output gap to the NRI, $\beta _{r}^{y}$, in
Eq. \ref{ygap}. Now we explain how we calibrated these parameters.

\subsubsection{The elasticity of the spread to Debt-GDP ratio \label%
{subsubsec: spread elasticity}}

The parameter $\beta _{lev}^{\Delta }$ in Eq. \ref{spread leverage} captures
the degree of the financial friction in the model. If $\beta _{lev}^{\Delta
}=0$, we return to the standard model in which the volume of debt does not
matter. In Appendix \ref{app: empirical evidence}, we empirically validate
the existence of such a financial friction in Israel. Specifically, we find
a positive relationship between the interest rate spread and debt to GDP
ratio in the Israeli mortgage market.

We are aware that the estimated elasticity we found also reflects the
macroprudential steps applied in Israel during the last decade (see \cite%
{Benchimol_et_al_2021}), therefore the estimated parameter\ is probably
upward biased regarding the pure parameter of the banking system. Thus, in
Appendix \ref{app: empirical evidence} we apply several tests by considering
the macroprudential steps. We find that even under accounting these
macroprudential steps, the basic parameter is still positive and
significant, and we find some support for the upward biasedness of the
estimated parameter.

More specifically, we estimate the following regression (in quarterly sample
2004:Q1-2021:Q3):%
\begin{equation}
spread_{t}^{H}=c+\beta Lev_{t}^{H}+\alpha _{0}\pi _{t}^{H}+\alpha _{1}\pi
_{t-1}^{H}+\alpha _{2}\pi _{t-2}^{H}+\alpha _{3}\pi _{t-3}^{H}+u_{t}
\label{reg spread lev}
\end{equation}%
Where the debt to GDP ratio defined as $Lev_{t}^{H}=\frac{B_{t}^{H}}{Y_{t}}$%
, $B_{t}^{H}$ is the stock of mortgages and $Y_{t}$ is the GDP (with
seasonal adjustment), $\pi _{t}^{H}$\ \ is housing prices. We choose this
specification to be close to our model Eq. \ref{spread leverage} and \cite%
{Benigno_et_al_2020}, where the spread is determined by the leverage in the
same period. Our purpose is to obtain an adequate estimator of the parameter 
$\beta $\ for model calibration, and not to obtain the best fit of Eq. \ref%
{reg spread lev} with respect to $R^{2}$. The inclusion of housing prices $%
\pi _{t}^{H}$\ is expected to capture a collateral effect on the spread, and
it is very significant in all regressions examined, while it helps in
identification of the true supply-side shocks to the spread\ $u_{t}$.

Under some assumption in the non-mortgage credit market, elaborated in
Appendix \ref{app: deriving elasticity}, we calibrate the model-consistent
parameter regarding to the total household credit market, $\beta
_{lev}^{\Delta }$ to be $0.031$. This parameter is close to the parameter in 
\cite{Curdia_Woodford_2010} (of $0.1$), but much higher than obtained in 
\cite{Benigno_et_al_2020} for the US (of $0.0078$).

\subsubsection{The elasticity of the NRI to spread}

\cite{Benigno_et_al_2020} obtained that the elasticity of the NRI to spread
for the US is $-2.5$. We assert that for a SOE, this elasticity should be
lower in magnitude, due to the possibility of the savers to substitute part
of domestic saving with foreign saving (they face higher transaction costs
when investing abroad). Here the NRI rate will decline when a positive shock
to the spread hits the borrowers, but this decline will be moderate compared
to the closed economy. Based on these considerations, we calibrated the
elasticity of the NRI to the spread to $-0.5$ ($\alpha _{\Delta }^{NRI}=0.5$%
).

Concerning the NRI in Eq. \ref{NRI}, we also modeled it as a function of
expected growth of domestic potential output and the expected growth of
actual output abroad following \cite{CGG_2002}. The estimated elasticities
for Israel of both variables were taken from \cite{Ilek_Segal_2022}. Another
justification for a lower parameter here compared to \cite%
{Benigno_et_al_2020} is that here the effect is on the output gap and not
only on aggregate consumption as in \cite{Benigno_et_al_2020}.

\subsubsection{The elasticity of the output gap to NRI}

The elasticity of the output gap with respect to NRI (parameter $\beta
_{r}^{y}$ in Eq. \ref{ygap}) is equal to the elasticity of the output gap
with respect to the short real interest rate of the CB. Several estimates of
this parameter exist for Israel. \cite{Chen_Zion_2021} obtained $\beta
_{r}^{y}$ of $-0.02$, which is falls far below all the others. \cite%
{Ilek_Segal_2022} estimate is $-0.1$, \cite{Benchimol_2016} estimate is $%
-0.5 $, \cite{Argov_Elkayam_2009} estimate lies between $-0.42$ and $-0.82$,
and \cite{argov2012moise} estimate is close to $-0.1$.\footnote{%
This model is presented in a non-linear form, therefore it is hard to obtain
this elasticity directly. We applied the following procedure to derive it
indirectly. In the first stage, we generated 1,000 observation using
stochastic simulation of the DSGE model. In the second stage, we regressed
the output gap on real interest rate, and other explanatory variables like
specification of IS equation in \cite{Laxton_et_al2006}. We obtained
estimated elasticity of the output gap to the real interest rate between
-0.1 and -0.15.} We adopt a conservative approach in calibrating parameter $%
\beta _{r}^{y}$ in Eq. \ref{ygap}, setting it to $-0.1$, which is a left
tail of all estimates for Israel, except the estimate in \cite%
{Chen_Zion_2021}.

\subsubsection{Borrowers' debt aversion}

Concerning other parameters in the financial block, the highest uncertainty
is about the parameter $v$ in Eq. \ref{Euler_borrowers}, that economically
reflects the degree of aversion to deviations of the leverage from its SS
value. Low values of $v$ induce high volatility of debt and leverage and
vice versa. \cite{Benigno_et_al_2020} set $v=0.025$ for the US. No estimates
or any assessments exist for Israel. One way to calibrate $v$ is to match
debt volatility from the model to the data. Measuring debt gap from the data
using HP filter results in SE of about $4\%$. To obtain this volatility,
high $v$ is required in the model, $v=1.25$. However, HP filter is a very
conservative approach of deriving gaps, because the HP trend closely tracks
the actual data by construction (the standard assumption of smoothing
parameter is $\lambda =1,600$). In contrast, deriving gaps from linear trend
induces remarkably high volatility of debt gap (under linear trend $\lambda $
is very big). Measuring debt gap from the data using linear trend results in
SE of $13\%$. To obtain such SE from the model, low parameter of $v$ is
needed as in \cite{Benigno_et_al_2020}, of about $v=0.0225$. Given the\ very
high uncertainty regarding to true volatility of debt gap in the data, we
examine two polar cases of this parameter, baseline case with $v=0.0225$ as
in \cite{Benigno_et_al_2020} and alternative case with $v=1.25$ which is
higher by $50$.

\subsubsection{Other parameters}

The parameter $\alpha _{cb}^{NRI}$ in Eq. \ref{NRI} is calibrated to $-0.4$,
capturing the share of borrowers in the population (according to \cite%
{Shami_2019}). Parameters in Eq. \ref{budget_b} are obtained from
linearization of the equation. Finally, we calibrate $\beta _{r}^{c^{b}}$ to
be $5$ times bigger than $\beta _{r}^{y},$ economically meaning that the
consumption of the borrowers is more sensitive to the interest rate than the
lenders' (as in \cite{Curdia_Woodford_2016}).

\begin{table}
\caption{\textbf{The parameters of the financial block of the model}}
\label{table:calibration}

\begin{tabular}{llll}
\hline\hline
$\text{Parameter}$ &  & $\text{Value}$ & $\text{Source or Target}$ \\ \hline
$\text{Elasticity of the Spread }$ &  &  &  \\ 
$\ \ \text{to Debt-GDP}$ & $\beta _{lev}^{\Delta }$ & $0.031$ & Appendix \ref%
{app: deriving elasticity} \\ 
$\text{Elasticity of the NRI to}$ &  &  &  \\ 
$\ \ \ $1. $\text{Spread}$ & $\alpha _{\Delta }^{NRI}$ & $0.5$ & \cite%
{Benigno_et_al_2020} \\ 
&  &  & and authors' considerations \\ 
\ \ 2. Domestic expected &  &  &  \\ 
\ \ \ \ \ \ potential growth & $\alpha _{g}^{NRI}$ & $0.4$ & \cite%
{Ilek_Segal_2022} \\ 
&  &  &  \\ 
\ \ 3. Actual growth abroad & $\alpha _{g^{\ast }}^{NRI}$ & $0.6$ & \cite%
{Ilek_Segal_2022} \\ 
&  &  &  \\ 
\ 4. Preference shock & $\alpha _{cb}^{NRI}$ & $0.4$ & Share of borrowers \\ 
&  &  &  \\ 
$\text{Elasticity of the Output Gap}$ &  &  &  \\ 
$\text{ \ \ to NRI}$ & $\beta _{r}^{y}$ & $-0.1$ & \cite{argov2012moise}, \\ 
&  &  & \cite{Ilek_Segal_2022} \\ 
Euler equation of borrowers &  &  &  \\ 
\  & $v$ & $0.0225$-$1.25$ & \cite{Benigno_et_al_2020} \\ 
&  &  & and authors considerations \\ 
& $\beta _{r}^{c^{b}}$ & $\beta _{r}^{y}\times 5$ & \cite%
{Curdia_Woodford_2016} \\ 
\  &  &  & and authors considerations \\ \hline
\end{tabular}

\end{table}%

\section{Implications\ of Financial Friction and Macroprudential Policy\label%
{Section: Implication of FF}}

In this section, we show the implications of financial friction and
macroprudential policy on the economy. To that end, we compare the responses
of the main variables to various shocks in a model under three cases, as
described below:

\begin{itemize}
\item \textbf{A model without financial friction}. In this case, commercial
banks do not take into account the possibility of default and don't penalize
borrowers for higher leverage. That means that the spread for the borrowers
does not depend on their leverage ratio ($\beta _{lev}^{\Delta }=0$ in Eq. %
\ref{spread leverage}) and it is completely exogenous.

\item \textbf{A model with financial friction}. The banks penalize borrowers
for higher debt, as a result of internal risk management, so they require
higher credit spread for higher leverage ratio ($\beta _{lev}^{\Delta
}=0.031 $, as under the basic calibration, see Table 
\ref{table:calibration}%
).

\item \textbf{A model with macroprudential policy}. Here the banks
influenced significantly by regulation of macroprudential policy, resulting
in a higher credit spread sensitivity to borrowers' leverage ($\beta
_{lev}^{\Delta }=0.1>0.031$). As discussed earlier, in Eq. \ref{spread
leverage}, macroprudential policy should try to prevent excess borrowing.
One possible tool is to impose stringent regulations on the borrowers, such
as maximum level of leverage (LTV) or maximum payment-to-income ratio (PTI).
Another common tool is by controlling the commercial banks' capital
requirements, which can make borrowing more costly. However, we do not
analyze here how each tool affects the spread sensitivity to borrowers'
leverage, $\beta _{lev}^{\Delta }$. Our purpose is to illustrate how an
increase in this sensitivity affects the economy.
\end{itemize}

\subsection{Implication of financial friction and macroprudential policy
under monetary policy shock}

\begin{figure}
\caption{\textbf{IRF to positive monetary policy shock}}
\includegraphics[width=\linewidth]{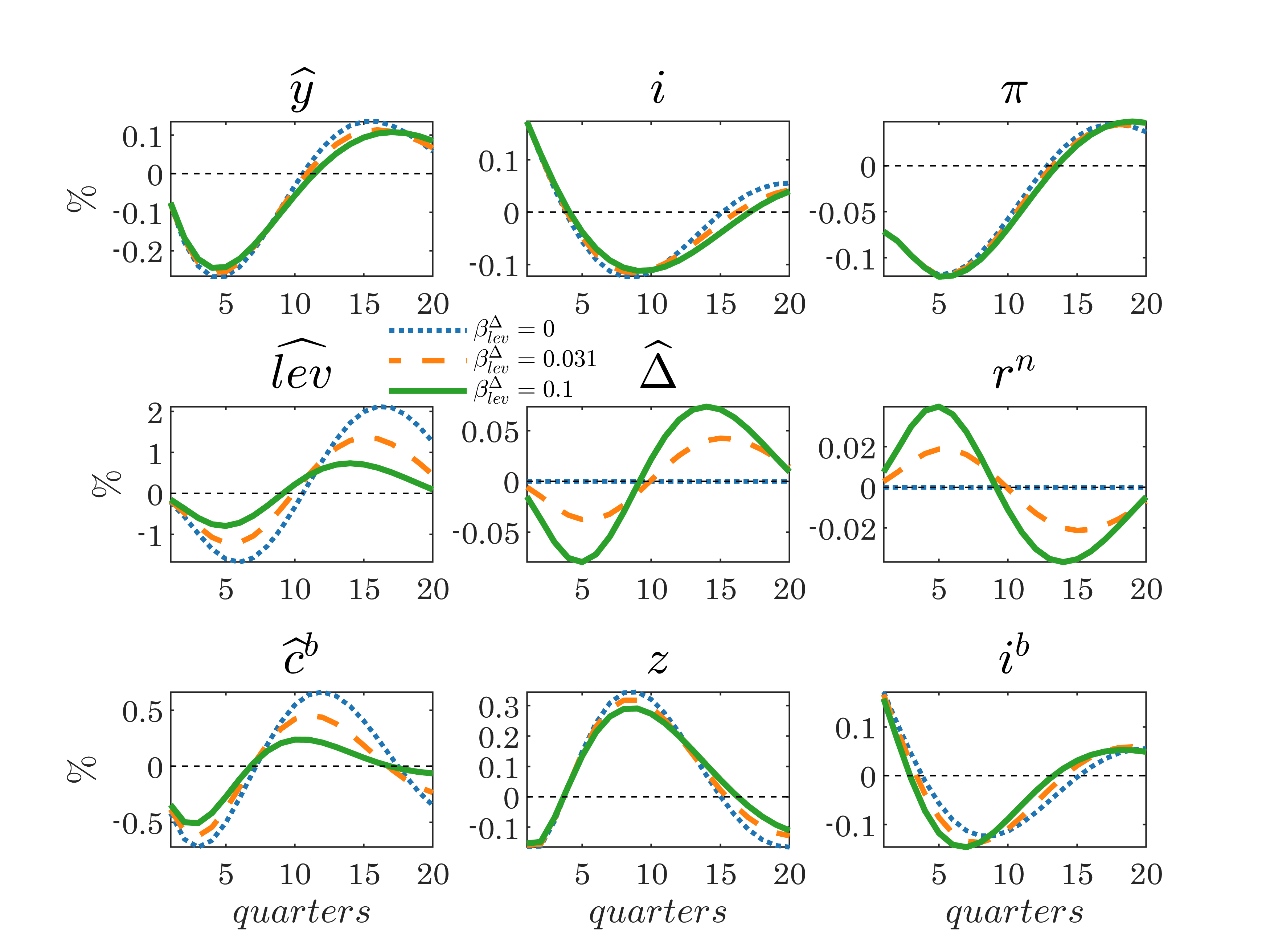}
\label{fig: IRF monetary policy}
{\footnotesize
\textit{Notes:} Shock of 1 standard deviation. Variables as deviation from steady state: output gap, inflation, interest rate,
leverage, credit spread, borowers consumption, real exchange rate and borowers' interest rate.
Blue dotted line: represents the case without financial friction. 
Orange line: represents the baseline case with financial friction, but without macroprudential policy 
(Elasticity  of spread to leverage is 0.031).
Green line: represents the case with macroprudential policy.
(elasticity  of spread to leverage is 0.1).
All simulations are under specification of low borowers' aversion to leverage, v=0.0225
}
\end{figure}

Fig. 
\ref{fig: IRF monetary policy}
shows the impulse response function (henceforth, IRF) to monetary policy
shock of nine variables - output gap, central bank interest rate, inflation,
leverage ratio, credit spread, natural rate of interest, borrowers'
consumption, real exchange rate and borrowers' risky interest rate. The blue
dotted line presents the case without financial friction, $\beta
_{lev}^{\Delta }=0$. The orange line represents the baseline case with
financial friction, $\beta _{lev}^{\Delta }=0.031$, but without distinctive
macroprudential policy. Finally, the green line represents the case with
tight macroprudential policy\textbf{\ }describe above, result in total
elasticity of $\beta _{lev}=0.1$.

The contractionary monetary policy shock reduces consumption demand through
the financial decision channel, since higher interest rate drives all
households -- borrowers and savers -- to reduce debt or increase saving,
accordingly. As a result, we get a decline in output and inflation. We can
see that under monetary policy shock the role of the financial friction and
macroprudential policy is not significant, and we obtain quite a similar
dynamic of the main variables -- output and inflation -- in all three cases.

In the case without financial friction (blue dotted line), the borrowers'
spread is not influenced by their leverage level, and it is constant. While
in the case with financial friction (orange line), the reduction in
borrowers' leverage gives a relief in the credit conditions and we obtain a
small decline in the credit spread. This has a minor positive contribution
to borrowers' demand. Finally, under macroprudential policy, we get much
higher reduction in the spread, with lower decline in leverage and
consumption of borrowers. This behavior of a slightly smoother output path
with higher financial friction is elaborated on the next section.

\subsubsection{\textbf{Financial Accelerator or Financial Decelerator?\label%
{subsubsection: Financial Accelerator or Financial Ddecelerator}}}

The sign of the response of credit spread to monetary shock is unequivocal
in the literature, even when considering the same country and the same
methodology. For example, \cite{Brzoza_Brzezina_et_al_2013} show, based on a
VAR model for the US economy, that a positive monetary shock reduces spread,
at least for one year. In contrast, \cite{Cesa_Bianchi_Anderson_2020} and 
\cite{Gertler_Karadi_2015} also consider the US economy but show that
spreads rise.

The main result we obtained previously is that a positive monetary shock
reduces leverage, and therefore reduces the spread. The reduction in the
spread partially mitigates the negative effect of the monetary shock on
consumption of borrowers and output. So, the financial friction in our model
calibration generates a \textbf{financial decelerator} rather than financial
accelerator. We will now elaborate on this result, starting by referring to
the literature, where one can find that there are few types of financial
frictions modeling approaches (\cite{Brzoza_Brzezina_et_al_2013} among
others).

Our point is that for all type of models, the spread increases with
leverage, but the leverage is defined differently, therefore its reaction to
shocks may be different. This is the reason why in different models the
impact of shocks is different -- it amplifies or mitigates with respect to
the case without financial friction. The reaction is depends critically on
the question of whether the leverage increases or decreases in response to
shocks -- which depends on the relative reaction of leverage numerator
(debt) and leverage denominator (net worth, house value, etc.).Specifically:

\begin{enumerate}
\item In \cite{Bernanke_et_al_1999} the financial friction is reflected by
the (negative) dependence of the credit spread on borrowers' (firms') net
worth.\footnote{%
In \cite{Gertler_Karadi_2011} the financial spread depends on the lender's
net worth (financial intermediates).}\ In this framework the financial
friction induces that shocks are amplified; hence it is named a financial
accelerator. After a (positive) monetary shock, the leverage increases
because the value of firms/banks assets (leverage denominator) decreases
more than the volume of credit (leverage numerator). This induces increase
in credit spread and causes amplification of the monetary shock.

\item In \cite{Kiyotaki_Moore_1997} and \cite{Iacoviello_2005} the friction
is reflected in existence of collateral constraint on borrowing, so leverage
is defined as Loan to Value (LTV). In this framework, the financial friction
induces financial accelerator for some types of shocks and financial
decelerator for others.

\item In \cite{Curdia_Woodford_2016} and \cite{Benigno_et_al_2020} the
leverage denominator is the SS level of debt and output, respectively. So,
in this case, the leverage decreases as a result of (positive) monetary
shock, because debt decreases along with constant denominator in the
leverage.
\end{enumerate}

In our model, as we show previously, the denominator is a time-variant
output gap (which is related to households' income). As such, our model can
be viewed as a generalization of \cite{Curdia_Woodford_2016} and \cite%
{Benigno_et_al_2020}, thus can function as financial accelerator or
decelerator, depending on calibration. In our model calibration (Table 
\ref{table:calibration}%
) the debt decreases more than the output, therefore the leverage and the
spread decrease, and we get deceleration (which was validated empirically in
Section \ref{subsubsec: the case of Israel}). If under some parameterization
a positive monetary shock reduces output more than debt, then the leverage
will rise and consequently the spread will increase, which will cause
acceleration (see Appendix \ref{app: IRF Financial Accelerator}), but these
parameters are not reasonable for Israel.

\subsubsection{The case of Israel\label{subsubsec: the case of Israel}}

Now we will validate the existence of a financial decelerator in Israel
empirically. To that end, we assess the sign (and the size) of elasticity of
the spread to monetary shock. If positive monetary shock induces higher
(lower) spread, it indicates existence of financial accelerator
(decelerator). To evaluate the sign of the effect of monetary shock on
spread in Israeli mortgage market we estimate the following regression:%
\begin{equation}
{\Large spread}_{t}^{H}{\Large =\alpha \epsilon }_{t}^{mon}+\beta
X_{t-1}+\gamma Z_{t}+u_{t}  \label{spread_dec}
\end{equation}%
where $\epsilon _{t}^{mon}$\ is a monetary shock based on projections for
the Bank of Israel (henceforth, BOI) interest rate by the professional
forecasters in Israel. $X_{t}$\ is a vector of explanatory variables
containing the leverage (debt over GDP), changes in the houses prices,
changes in the unemployment rate and the spread. Vector $Z_{t}$\ contains
exogenous variables including a dummy variable representing macroprudential
policy measures of the BOI in the housing market (based on calculation of 
\cite{Benchimol_et_al_2021}), and the VIX index in the US representing
uncertainty. $u_{t}$\ are residuals.

When the spread in Eq. \ref{spread_dec} is defined relative to the real rate
of commercial banks, we obtain $\alpha =-0.74$\ ($pvalue=0.045$), and when
it is defined relative to real return of government bonds we obtain $\alpha
=-0.77$\ ($pvalue=0.06$). The negative sign of $\alpha $\ is robust to other
specifications of Eq. \ref{spread_dec}, while its size can exceed -1 in some
cases.\footnote{%
In \cite{Cesa_Bianchi_Anderson_2020} the estimated elasticity also exceeds 1
for some tests. However, in contrast to our result the elasticity in \cite%
{Cesa_Bianchi_Anderson_2020} is positive.} So, we obtained empirical support
that the financial friction in the mortgage credit market in Israel induces
financial decelerator, similar to \cite{Curdia_Woodford_2016} and \cite%
{Benigno_et_al_2020}. This result gives support for the IRF of credit spread
to positive monetary shock in Fig. 
\ref{fig: IRF monetary policy}%
.

\subsection{Implication of financial friction and macroprudential policy
under credit supply shock\label{subsection: Implication under credit supply
shock}}

\begin{figure}
\caption{\textbf{IRF to negative credit supply shock}}
\includegraphics[width=\linewidth]{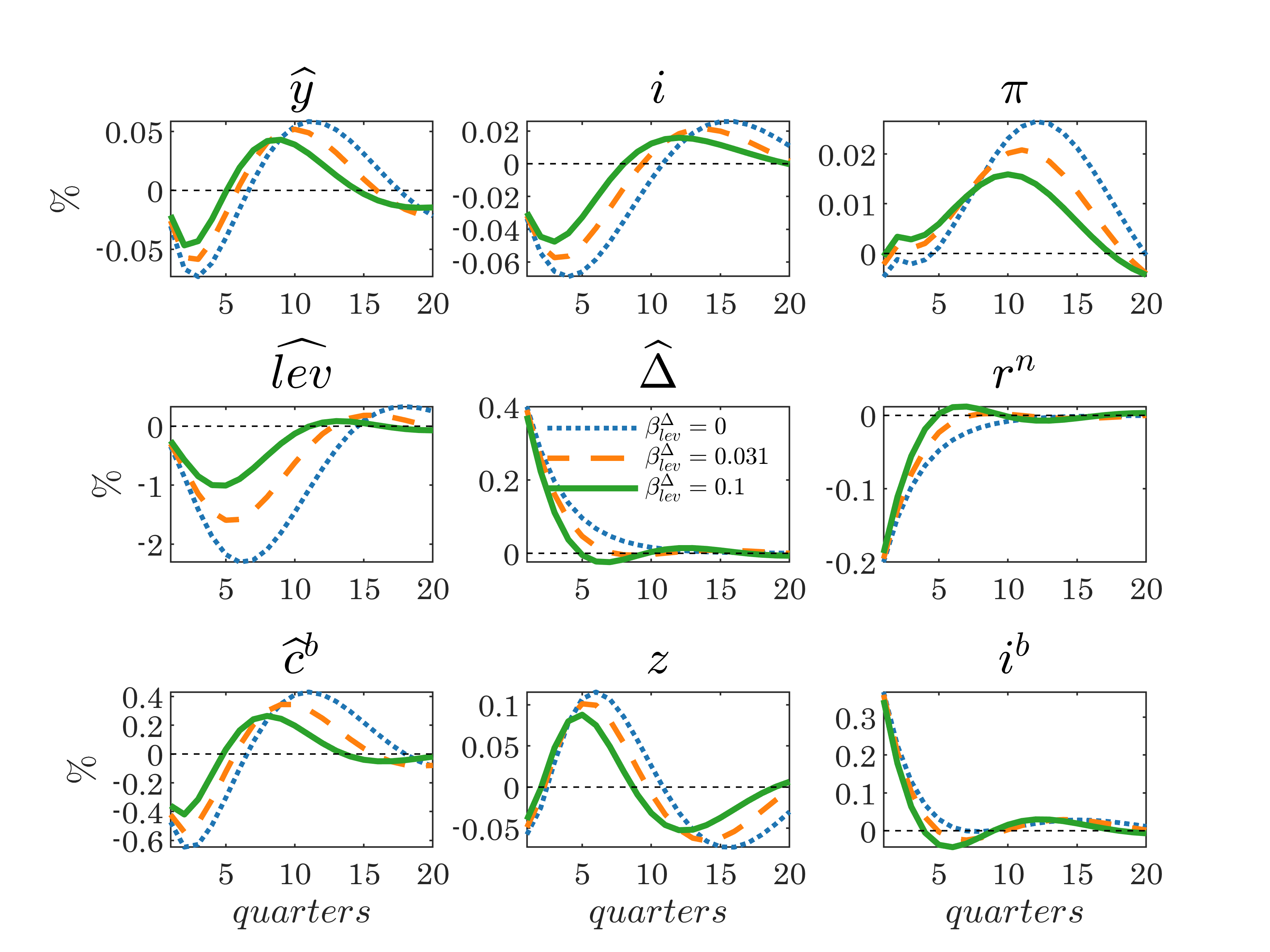}
\label{fig: IRF spread}
{\footnotesize
\textit{Notes:} Positive shock to spread of 1 standard deviation.Variables as deviation from steady state: output gap, inflation, interest rate,
leverage, credit spread, borowers consumption, real exchange rate and borowers' interest rate.
Blue dotted line: represents the case without financial friction. 
Orange line: represents the baseline case with financial friction, but without macroprudential policy 
(Elasticity  of spread to leverage is 0.031).
Green line: represents the case with macroprudential policy.
(elasticity  of spread to leverage is 0.1).
All simulations are under specification of low borowers aversion to leverage, v=0.0225
}
\end{figure}%

Fig. 
\ref{fig: IRF spread}
shows the IRF to positive credit spread shock which represents a negative
credit supply shock. In the case \textit{without} financial friction ($\beta
_{lev}^{\Delta }=0$, blue line) the spread is exogenous and independent of
the borrowers' leverage ratio. Since borrowers have a higher effective
interest rate, their debt is more expensive, and they prefer to deleverage.
This requires reduction in the borrowers' expenditure for consumption.
Simultaneously, the increase in the spread induces a decline in the NRI and
a reduction in the output gap. In the case \textit{with} financial friction
(orange line) the spread has an endogenous part, so as the borrowers
deleverage, it helps to reduce their credit spread. In turn they keep
deleveraging and reduce consumption, but much softer than in the first case.
Under macroprudential policy (ex-post policy) deleveraging process
contributes much more to the reduction of the spread, and thus the
deleveraging process is much more moderate, and we have a smoother path for
all variables, which is good policy outcome.

Note that a negative credit shock (which is a mirror image of the IRF's
described above in Fig. 
\ref{fig: IRF spread}%
) represents credit market easing which results in economic expansion. In
these circumstances, the macroprudential policy (ex-ante policy) acts to
limit excess leverage buildup and reduces the likelihood of a financial
default.

Looking at the second-year response (quarters 5 to 8)\ we can see that
borrowers' consumption and aggregate activity are high. This can be
explained by the real depreciation along the first year (increasing $z_{t}$)
which supports modest inflation in the first year and accumulates to high
yearly inflation. Due to high inflationary inertia, as can be seen in Eq. %
\ref{philips}, we get high inflation rate along the second year. This
inflationary path with smoothing (inertial) interest rate policy has the
consequences of lowering the real interest rate and has expansionary impact
of high activity. In other words, the monetary policy is not contractionary
enough (due to smoothing)\ and unintentionally causes an expansionary
effect. This mechanism based on the interaction between spread (financial
market) and real exchange rate does not exist in close economic models.

\subsection{Implication of financial friction and macroprudential policy
under credit demand shock\label{subsection: Implication under credit demand
shock}}

\begin{figure}
\caption{\textbf{IRF of positive credit demand shock (borrowers with low aversion to leverage)}}
\includegraphics[width=\linewidth]{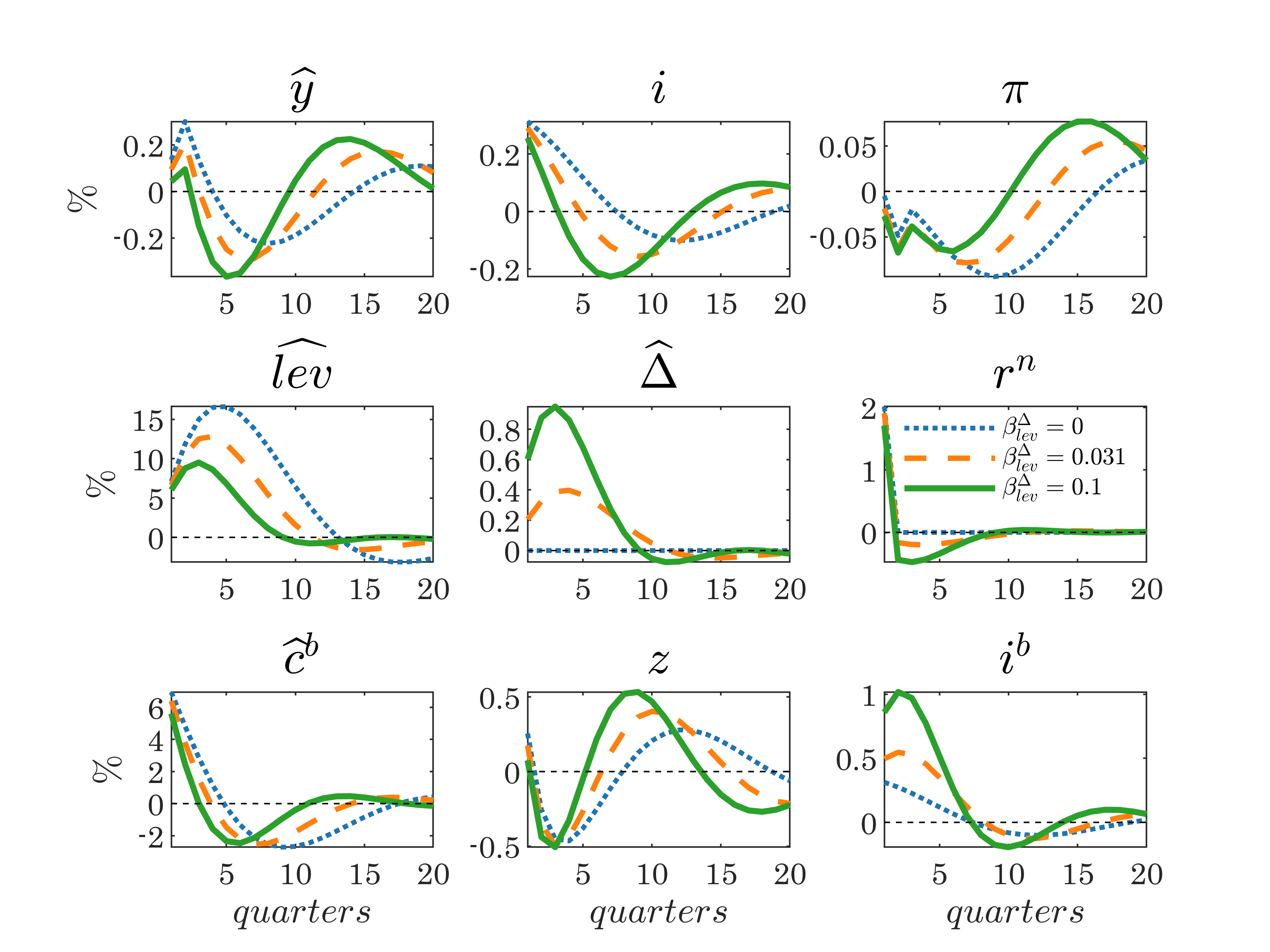}
\label{fig: IRF to prefernce shock}
{\footnotesize
\textit{Notes:} Positive preferences shock  of 1 standard deviation. Variables as deviation from steady state: output gap, inflation, interest rate,
leverage, credit spread, borowers consumption, real exchange rate and borowers' interest rate.
Blue dotted line: represents the case without financial friction. 
Orange line: represents the baseline case with financial friction, but without macroprudential policy 
(Elasticity  of spread to leverage is 0.031).
Green line: represents the case with macroprudential policy.
(elasticity  of spread to leverage is 0.1).
All simulations are under specification of low borowers' aversion to leverage, v=0.0225
}
\end{figure}%

In Fig. 
\ref{fig: IRF to prefernce shock}
we induce a positive preference shock only to the borrowers ($\varepsilon
_{t}^{cb}$), which is translated into a higher demand for borrowers
consumption, as can be seen in Eq. \ref{Euler_borrowers}, and thus a higher
demand for credit (see Eq. \ref{budget_b}). Accordingly, we call this shock
a credit demand shock. On impact, this shock increases the NRI (see Eq. \ref%
{NRI}), which induces output gap increase, which is consistent with the
borrowers' consumption increase. This is pronounced in the case without
financial friction ($\beta _{lev}^{\Delta }=0$) as the leverage rises
sharply, while the spread does not move. While in the case with financial
friction, especially when the macroprudential policy is at place ($\beta
_{lev}^{\Delta }=0.1$), after credit demand shock the leverage increases,
and the spread increases dramatically, which in turn lower NRI and a have
negative effect on aggregate activity. In this case, the substantial
increase in credit spread mitigates the excess borrowers demand and
leveraging, by increasing borrowing costs. In summary, there is \textit{%
trade-off} in implementation of macroprudential policy, since as
macroprudential policy is more solid (higher $\beta _{lev}$) the spread
increases significantly, and thus the policy succeeds in mitigating increase
in the leverage, which is \textit{beneficial} for financial stability
(ex-ante prevention policy). But this is at the\textit{\ cost} of higher
spread which lowers the\ NRI and deteriorates aggregate real activity, which
is a bad outcome. We will discuss this further in Section \ref{Section:
Monetary policy analysis}.

\begin{figure}
\caption{\textbf{IRF of positive preferences shock (borrowers with high aversion to leverage)}}
\includegraphics[width=\linewidth]{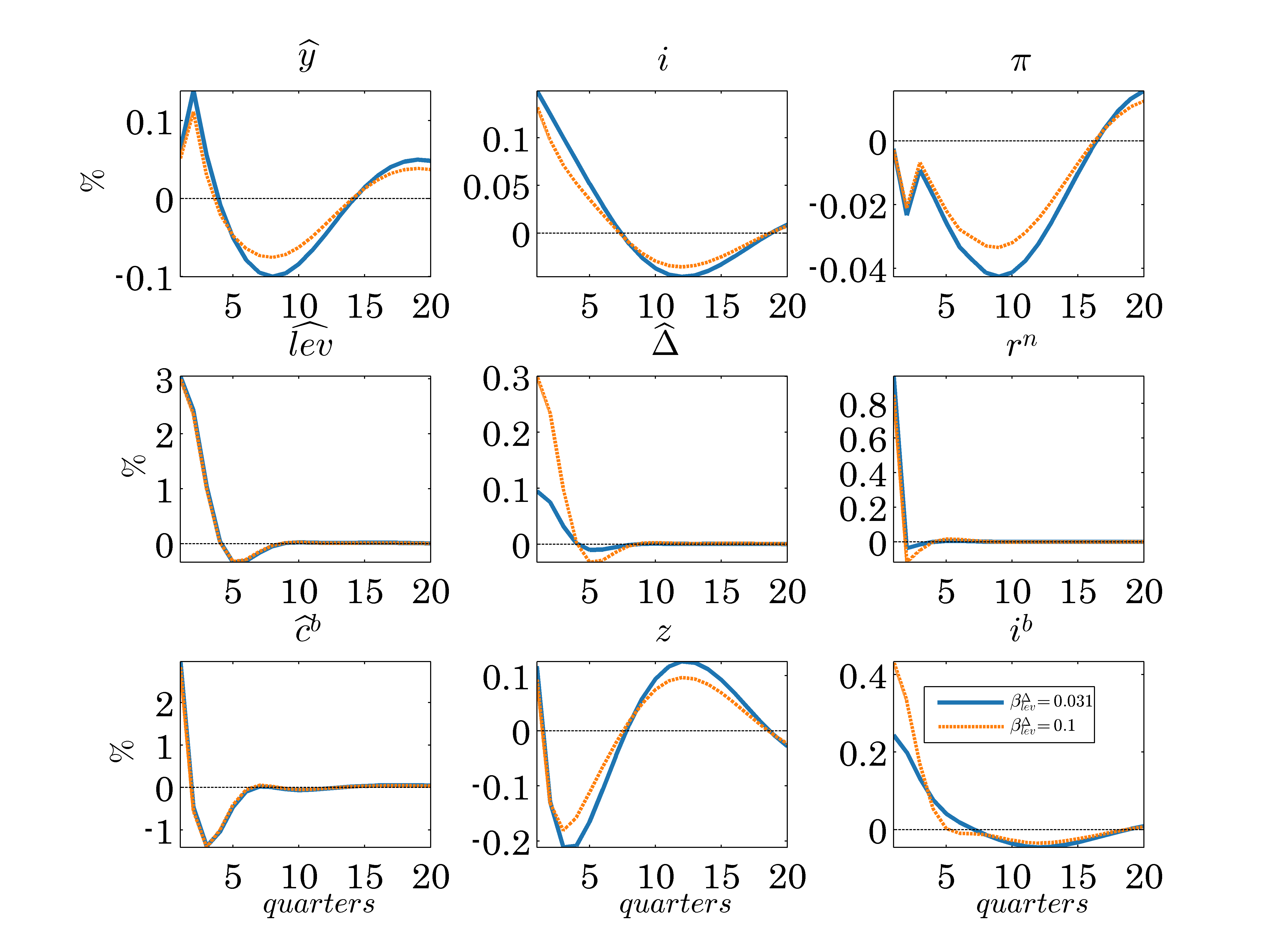}
\label{fig: IRF prefernce shock high aversion}
{\footnotesize
\textit{Notes:} Variables as deviation from steady state: output gap, inflation, interest rate,
leverage, credit spread, borowers consumption, real exchange rate and borowers' interest rate.
Blue dotted line: represents the case without financial friction. 
Orange line: represents the baseline case with financial friction, but without macroprudential policy 
(Elasticity  of spread to leverage is 0.031).
Green line: represents the case with macroprudential policy.
(elasticity  of spread to leverage is 0.1).
All simulations are under specification of high borowers' aversion to leverage, v=1.25
}
\end{figure}

\textbf{Debt Aversion. }The effectiveness of macroprudential policy depends
very much on the degree of aversion of the borrowers with respect to
deviation of debt from its normal value (see Eq. \ref{Euler_borrowers}). To
show this, Fig. 
\ref{fig: IRF to prefernce shock}
and Fig. 
\ref{fig: IRF prefernce shock high aversion}
present the IRFs with low ($v=0.0225$) and high ($v=1.25$) debt aversion
parameter. It can be seen that for the same demand shock, households with
low risk aversion are much more affected by the macroprudential policy than
households with high degree of aversion. This is reflected in significant
reduction of the leverage to increase in the parameter $\beta _{lev}^{\Delta
}$ in a case with low risk aversion (Fig. 
\ref{fig: IRF to prefernce shock}%
). However, in a case with high risk aversion, Fig. 
\ref{fig: IRF prefernce shock high aversion}%
, we can see it just barely react to the macroprudential policy. The reason
for this result is that households with high aversion are very conservative
from the beginning, and they are reluctant to deviate from normal leverage.
This can be seen in moderate increase of their leverage by 3\% in Fig. 
\ref{fig: IRF prefernce shock high aversion}
as opposed to 15\% in Fig. 
\ref{fig: IRF to prefernce shock}%
. Therefore, making the borrowing costs more expensive (higher $\beta
_{lev}^{\Delta }$) does not matter very much for borrowers with high
aversion. In contrast, households that do not care very much about the
volume of their debt from the beginning are more sensitive to the cost of
borrowing, making macroprudential policy more efficient.

\section{Monetary policy analysis\label{Section: Monetary policy analysis}}

In this section, we examine the implications of financial friction in the
household sector credit market on monetary policy. Our goal is not to derive
an optimal monetary policy rule, but to analyze quantitatively the
consequences on the economy when the monetary policy ignores the financial
sector and responds only to standard variables, such as inflation and real
activity.

\cite{Curdia_Woodford_2010} (and \cite{Curdia_Woodford_2016}) examine
optimal reaction of monetary policy to credit spread. They show that if the
financial shock is dominant in the economy, a (negative) reaction to spread
(LAW policy) could be beneficial but the optimal size of the response
depends very much on the persistence of the financial shock. They also show
that reaction to financial spread is also beneficial to certain other types
of shocks but not for all types. Moreover, the size and even their sign
depend on the type of shock and on the degree of its persistence. So, it is
hard to provide a robust recommendation on how optimally monetary policy
should react to financial spread. This result is in line with \cite%
{McCulley_Toloui_2008}, who proposed deducting the spread from the natural
rate of interest. In this sense, our policy rule, which negatively reacts to
the spread, is consistent with the above papers.

In our model, the financial spread is an integral part of the NRI. Thus, the
only possibility we will consider here is monetary policy (see Eq. \ref%
{policy_rule}) that react to the spread vs. monetary policy that not react
to the spread. It is important to note that all models' equations are the
same, and just the policy rule, Eq. \ref{policy_rule}, will be different by
taking misspecified NRI as we will elaborate below (Eq. \ref{NRI_wrong})
instead of the correct model NRI (Eq. \ref{NRI}). In details, in the
following exercise we compare the IRF of the main economic variables to
shocks under two policy rules: the first rule reacts to correctly specified
NRI in Eq. \ref{NRI} which takes into account the credit spread and
preference shocks of the borrowers, in line with LAW policy (we will call it
"FI" -- Fully Informative). The alternative policy rule reacts to
misspecified NRI, which corresponds to non-LAW policy (we will call it "PI"
-- Partially Informative), and defined as 
\begin{equation}
r_{t}^{n,PI}=\alpha +\alpha _{g}^{NRI}g_{t+1}^{n}+\alpha
_{g^{w}}^{NRI}g_{t+1}^{w}.  \label{NRI_wrong}
\end{equation}%
The only difference between Eq. \ref{NRI_wrong} and Eq. \ref{NRI} is that
the former rule ignores the credit spread as well as preference shocks of
the borrowers. The gap between the correct and misspecified NRI (exploiting
Eq. \ref{spread leverage}) is

\begin{equation}
r_{t}^{n}-r_{t}^{n,PI}=\alpha _{\Delta }^{NRI}(\beta _{lev}^{\Delta }%
\widehat{lev}_{t}+\varepsilon _{t}^{\Delta })+\chi _{cb}^{NRI}\Delta
\varepsilon _{t+1}^{cb}  \label{NRI_gap}
\end{equation}%
Thus, it is easy to see that the difference between the IRFs stems mainly
and directly from two shocks, $\varepsilon _{t}^{\Delta }$ and $\Delta
\varepsilon _{t+1}^{cb}$. While other shocks also have some indirect impact
through effect on leverage ratio,\ $\widehat{lev}_{t}$, quantitatively the
effect of leverage in our model is tiny, because according to our
calibration, $\alpha _{\Delta }^{NRI}\cdot \beta _{lev}^{\Delta
}=\allowbreak 0.016$. So, any reasonable change in the leverage ratio due to
shocks in the economy has only a small impact on the difference between the
IRF's under "FI" and "PI" rules. It comes out that the IRFs under the two
policy rules are almost identical for all shocks except two, $\varepsilon
_{t}^{\Delta }$ and $\Delta \varepsilon _{t+1}^{cb}$. Fig's 
\ref{fig: IRF spread low aversion 2}
to 
\ref{fig:IRF preference shock high aversion 2}
present the IRF to these two shocks, where "R-Correct" corresponds to the
"FI" rule, whereas "R-Wrong" corresponds to the "PI" rule.

Below we consider two cases. In Case 1, the households have low aversion to
deviations of the leverage from its SS value ($v$ in Eq. \ref%
{Euler_borrowers} is low), and in Case 2 they have high aversion ($v$ is
high).

\subsection{Case 1: Households with low aversion to leverage deviations}

Fig. 
\ref{fig: IRF spread low aversion 2}
shows the IRF to a positive spread shock (of 0.4 p.p). Higher spread makes
debt more expensive, therefore the leverage ratio ($\widehat{lev}_{t}$)
declines by about 2 p.p (after 5 periods). The reduction in debt forces
borrowers to reduce consumption ($\widehat{c}_{t}^{b}$). Looking now at the
aggregate economy, the NRI declines by 0.2 p.p (see Eq. \ref{NRI}) pushing
the output gap downward (see Eq. \ref{ygap}). Under the "FI" rule, the CB
immediately cuts the interest rate following reduction in the NRI. This
"on-time" reaction of the CB mitigates the fall in the output gap and
inflation. It also mitigates the fall in the borrower's consumption. The
picture is different under the "PI" rule. The CB does not react to the
spread, because its perceived NRI is misspecified (see Eq. \ref{NRI_wrong}).
As a result, the output gap noticeably falls and inflation as well (compared
to the "FI" rule). The reaction of the CB is delayed and stronger compared
to the "FI" rule, because in the former case the policy reacts only to the
(noticeable) fall in the output gap and inflation. Overall, due to reaction
of the CB to the correct NRI, it manages to stabilize the economy more
efficiently than under incorrect NRI - a result which is reflected in more
moderate IRFs of output and inflation. If the Monetary Committee will decide
not to react to the spread, eventually it will be forced to react strongly
in the future.

\begin{figure}
\caption{\textbf{IRF to positive preference shock of the borrowers of 1 SE, with low aversion to leverage}}
\includegraphics[width=\linewidth]{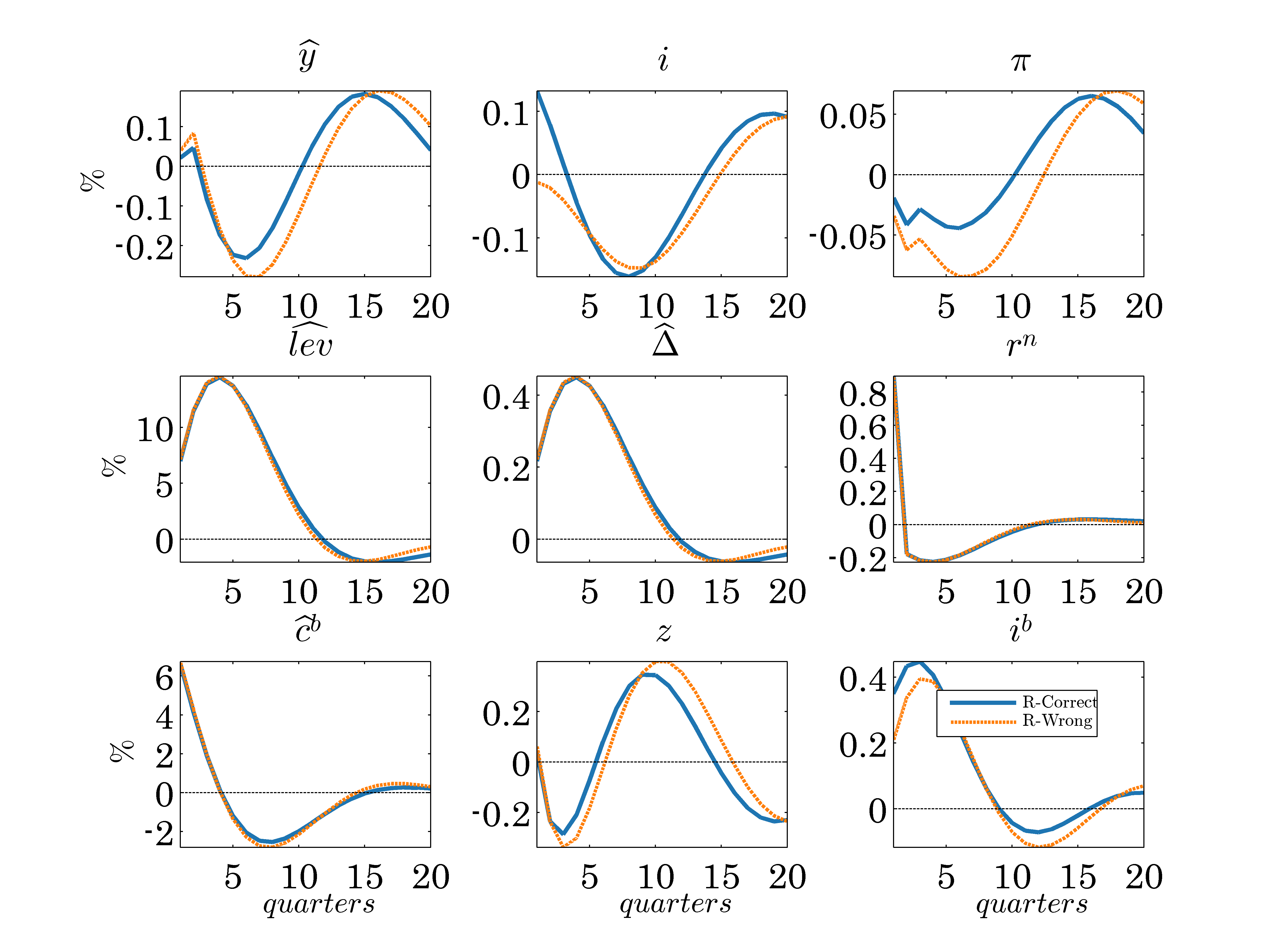}
\label{fig: IRF preference shock low aversion 2}
{\footnotesize
\textit{Notes:} R-Correct: CB reacts to correct NRI. R-Wrong: CB reacts to wrong NRI
low aversion to leverage, v=0.0225
}
\end{figure}

Fig. 
\ref{fig: IRF preference shock low aversion 2}
shows the IRF to a positive demand shock of the borrowers.\footnote{%
The SE of preference shock is taken from \cite{argov2012moise}.} The consumption
of the borrowers increases by 6\%, as well as their leverage ratio by about
10 p.p.. As a result, the spread increases by 0.2-0.4 p.p. (Fig. 
\ref{fig: IRF preference shock low aversion 2}%
). The preference shock also pushes the NRI upward (see Eq. \ref{NRI}) (note
that there is also indirect impact of the shock on the NRI through the
spread, but it is small). Overall, the deviations of output gap and of
inflation are slightly more moderate under "FI" than under "PI" rule, so the
policy is more effective. One can argue that the preference shocks of the
borrowers are hard to identify in practice, because they are unobserved.
Therefore, once these shocks are in place the assumption of the "FI" rule is
highly unreasonable. Note, however, that preference (demand) shocks are
reflected in the observed leverage ratio and debt volume in the same
direction (see Fig. 
\ref{fig: IRF preference shock low aversion 2}%
). In other words, the CB can optimally react to both financial spread and
leverage ratio and it is insufficient to react only to spread when credit
demand shocks are in place.

\subsection{Case 2: households with high aversion to leverage deviations}

\begin{figure}
\caption{\textbf{IRF to positive shock to spread of 1 SE, with low aversion to leverage}}
\includegraphics[width=\linewidth]{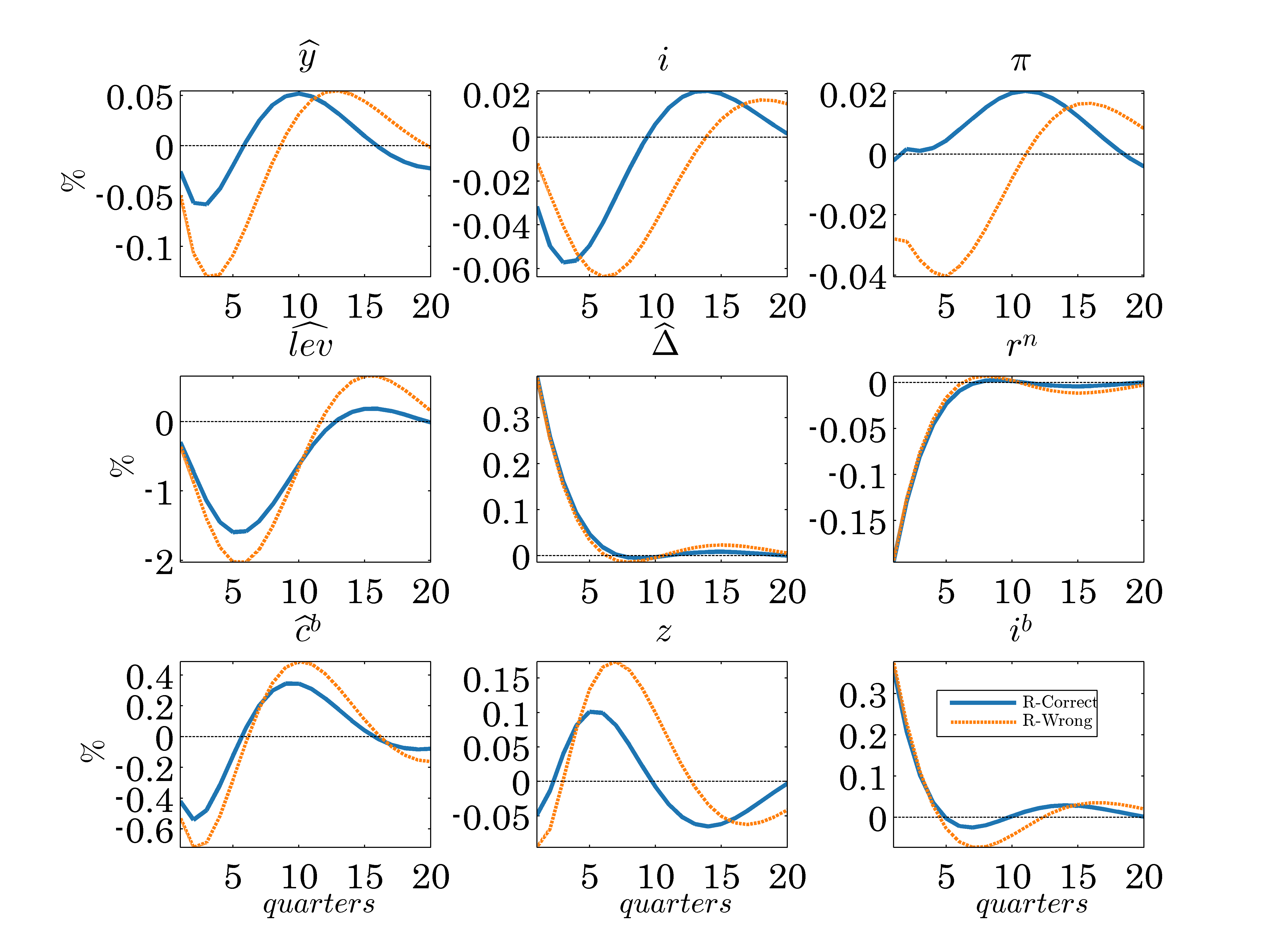}
\label{fig: IRF spread low aversion 2}
{\footnotesize 
\textit{Notes:} R-Correct: CB reacts to correct NRI. R-Wrong: CB reacts to wrong NRI. Low aversion to leverage is v=0.0225.
}
\end{figure}

\begin{figure}
\caption{\textbf{IRF to positive shock to spread of 1 SE, with high aversion to leverage}}
\includegraphics[width=\linewidth]{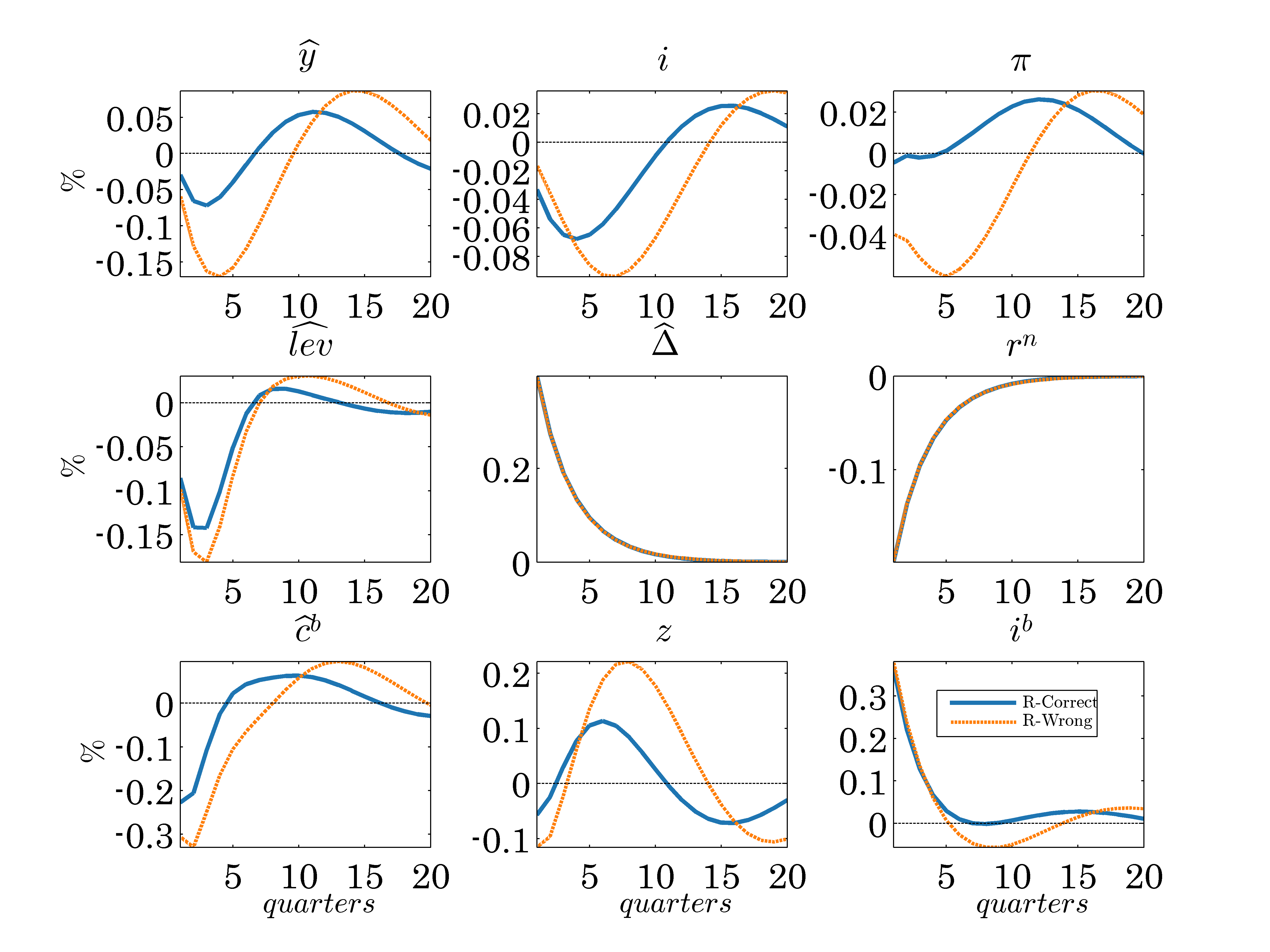}
\label{fig: IRF spread high aversion 2}
{\footnotesize
\textit{Notes:} R-Correct: CB reacts to correct NRI. R-Wrong: CB reacts to wrong NRI.
high aversion to leverage: v=1.25
}
\end{figure}

Fig. 
\ref{fig: IRF spread high aversion 2}
shows the IRF to the shock to interest rate spread when borrowers have high
aversion to debt deviations. The sensitivity of the leverage to the spread
shock here is much lower than in Fig. 
\ref{fig: IRF spread low aversion 2}%
, which is reflected in a lower decline in consumption. The IRF of the
output gap and inflation are similar to Fig. 
\ref{fig: IRF spread low aversion 2}%
, because these variables are affected by the NRI, which declines at a
similar rate as in Fig. 
\ref{fig: IRF spread low aversion 2}%
.

\begin{figure}
\caption{\textbf{IRF to positive preference shock of the borrowers of 1 SE, with high aversion to leverage}}
\includegraphics[width=\linewidth]{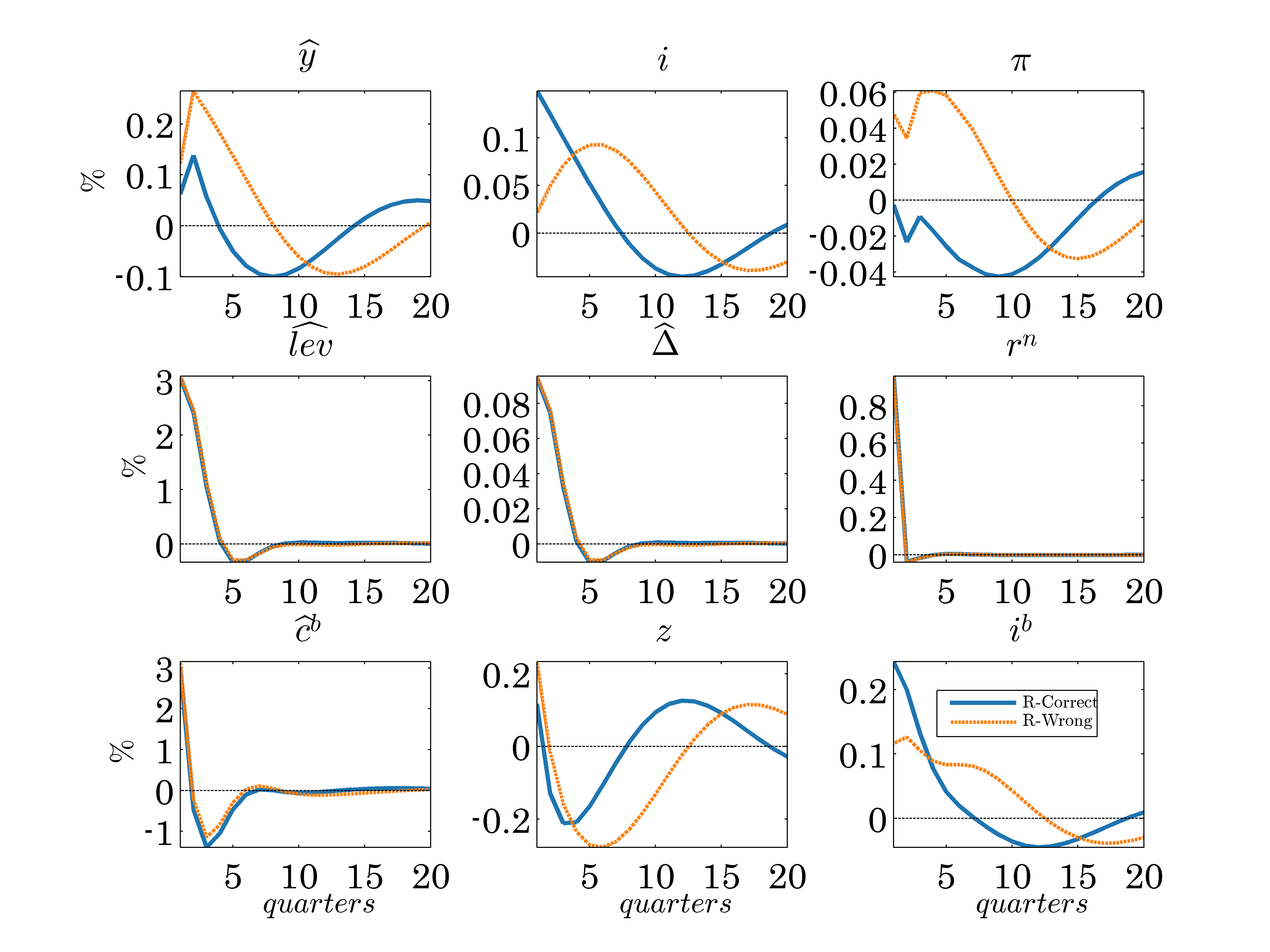}
\label{fig:IRF preference shock high aversion 2}
{\footnotesize
\textit{Notes:} R-Correct: CB reacts to correct NRI. R-Wrong: CB reacts to wrong NRI
high aversion to leverage: v=1.25
}
\end{figure}

Fig. 
\ref{fig:IRF preference shock high aversion 2}
shows the IRF to the preference shock. Households with high aversion to debt
increase their demand for consumption and credit only by 3\% (as opposed to
6\% and 15\%, respectively, in Case 1). As a result of moderate increase in
the leverage, the spread increases only by 0.1 p.p.. It is evident from Fig. 
\ref{fig:IRF preference shock high aversion 2}
that the under the "FI" rule the CB is more efficient in stabilizing
inflation and real activity.

\subsection{Loss function analysis}

The IRF's comparison shown above sheds first light on a loss function of the
CB when it ignores the financial sector. To make loss function analysis more
precise and formal, we make stochastic simulation with 10,000 periods and
calculate the loss function under two policy rules, "FI", and "PI". In the
simulations we activate only two shocks--credit demand and supply--because
as we explained previously, other shocks barely make any difference between
the IRFs under these two rules. We consider four versions of the loss
function as shown in Eq. \ref{loss}. Our purpose is to choose the loss
function that characterizes best the preference of the Monetary Committee
according to the BOI Law. The first version of the loss function (Version 1)
is based on the variances of inflation and output gap, which reflects two
main goals--price stability and stabilization of real activity. This loss
function was considered for Israel by \cite{Benchimol_2022}. We also
consider another version of the loss function (Version 2), which may also
match the preferences of the Monetary Committee in Israel. The only
difference is that the variance of the output gap is replaced by the
variance of \textit{changes} in the output gap--which is the gap between
actual and potential output growth. The third version (Version 3) also
includes the variance of \textit{changes} in the BOI interest rate. This
specification is consistent with \cite{Segal_2007}, who examined this loss
function for Israel in sample 1999-2007. Finally (Version 4) is like
(Version 3), but here the variance of \textit{changes} in the output gap is
included instead of variance of output gap.

Unfortunately, no up-to-date estimates for $\alpha $ and $\beta $ exist for
Israel in Eq. \ref{loss}. \cite{Segal_2007} considered a wide range of $%
\alpha $ between zero and 1.5 and an even larger range for $\beta ,$ between
zero and 16. \cite{Benchimol_2022} calibrated $\alpha $ between zero and 1
but he assumed $\beta =0$. We calibrate $\alpha =0.5$ (middle of range from 
\cite{Segal_2007} and \cite{Benchimol_2022}), and $\beta =6$ (relevant for
Versions 1 and 2 of the loss function). \cite{Segal_2007} showed that for $%
\beta =6$ (and $\alpha =0.5$) the derived interest rate replicates quite
well the actual interest rate in the sample 1999-2007, and this result is
robust for any $\beta <16$ and $0<\alpha <1.5$.

\begin{equation}
\begin{tabular}{ll}
$\text{Version 1}\text{: }$ & $var(\pi _{t}-\pi )+\alpha \cdot var(\hat{y}%
_{t})$ \\ 
$\text{Version 2}\text{: }$ & $var(\pi _{t}-\pi )+\alpha \cdot var(\Delta
y_{t}-g_{t}^{n})$ \\ 
$\text{Version 3}\text{: }$ & $var(\pi _{t}-\pi )+\alpha \cdot var(\hat{y}%
_{t})+\beta \cdot var(\Delta i_{t})$ \\ 
$\text{Version 4}\text{: }$ & $var(\pi _{t}-\pi )+\alpha \cdot var(\Delta
y_{t}-g_{t}^{n})+\beta \cdot var(\Delta i_{t})$%
\end{tabular}
\label{loss}
\end{equation}

Table 
\ref{table:loss}
reports the calculation of the loss function under the "FI" and "PI" rules
and the ratio between them. The loss function was calculated by assuming
moderate aversion of households to debt, that is, the parameter $v$ is equal
to $0.64$ (middle of range in Table 
\ref{table:loss}%
). It can be seen that under all versions of the loss function, under "PI"
rule the loss is always larger (the ratio lies between 1.13-3.12). The ratio
under Versions 3-4 is lower than under Versions 1-2, so the "FI" is less
beneficial that the "PI". The reason for that result is that under Versions
3-4 the change of the BOI interest rate is also included into the loss
function. Although the "FI" policy rule is more efficient in stabilizing
inflation and real activity, the BOI interest rate is more volatile because
it reacts to correct but more volatile NRI (see Eq. \ref{NRI} versus Eq. \ref%
{NRI_wrong}).

\begin{center}
\begin{table}
\caption{\textbf{The loss of the CB under Cases 1 and 2}}
\label{table:loss}%

\begin{equation*}
\begin{tabular}{|c|c|c|c|}
\hline
Loss & Case 1:FI & Case 2: PI & Ratio \\ \hline
Version 1 & 0.08 & 0.24 & 3.12 \\ \hline
Version 2 & 0.26 & 0.56 & 2.18 \\ \hline
Version 3 & 0.25 & 0.28 & 1.13 \\ \hline
Version 4 & 0.43 & 0.60 & 1.40 \\ \hline
\end{tabular}%
\end{equation*}

\end{table}%
\end{center}

\section{Conclusions\label{Section: Conclusions}}

There is still an open debate in the literature concerning the question of
whether central banks should react to financial variables, like spread or
credit, and if and how macroprudential policy can prevent future crises. We
built an empirical model for Israel which incorporates the credit market of
households, aiming to analyze the impacts of monetary policy and
macroprudential policy.

Investigating the stylized facts of Israeli households' credit market, we
find that the relationship between the credit spread and the debt leverage
is positive, but seems to be weak. This is reflected in low elasticity of
spread to LTV and to debt-GDP ratio. The upward continuous trend in the LTV
and in the debt-to-GDP ratio observed in recent years in Israel probably
indicate that the costs of borrowing are not high enough to mitigate this
trend. Another important fact is that the households in Israel hold only
domestic debt (the foreign debt is negligible).

Relying on this empirical evidence, we specify and calibrate a
semi-structural DSGE model for the Israeli economy with household's credit
market. As mentioned, since the households cannot borrow from abroad, the
credit spread is determined only by the domestic financial intermediares,
and also may be affected by macroprudential policy. If higher debt does not
increase the costs of borrowing very much, households have an incentive to
keep increasing their debt further. This boosts a probability of default in
the future and may have disastrous effects on the economy if default
materializes. Thus, in our model macroprudential policy aims to mitigate
excess borrowing by imposing regulations on the financial intermediares.
Such measures are eventually translated into a higher elasticity of credit
spread to credit leverage, mitigating incentive to borrow. The effectiveness
of this policy depends also on the degree of aversion of households to high
debt, since if households are conservative, they internally will not
over-borrow, and the policy effectiveness will be low. But if households
have low degree of aversion to high debt, such that they barely internalize
the effects of their borrowing decisions on credit spread and therefore face
a low effective interest rate in consumption decision, they may seek to
increase their borrowing very much. In this case macroprudential policy is
crucial and also effective. However, the direct cost of mitigating excess
borrowing is reflected in shrinking of real activity.

Analyzing the model, we provide the following findings. First, if monetary
policy does not react to credit spread, but reacts only to standard
variables, such as inflation and output gap, it loses the effectiveness to
achieve its main goals, which are stabilizing inflation and real activity.
Effectiveness deteriorates even more, if the central bank completely ignores
developments in the credit market, which can be induced, for example, by
shocks in preference of the borrowers. Second, as macroprudential policy may
increase the elasticity of credit spread to households' leverage, it can
mitigate or even prevent over-borrowing and reduce deleveraging crises risk.
In addition, in case of demand weakness, and deleveraging, this policy may
contribute to expansionary efforts, due to corresponding reduction in credit
spread.

This paper focuses on the qualitative analysis of a model with households'
credit and policy implications, adjusted for the Israeli economy. Even
though the model has been adjusted and calibrated to fit the Israeli data,
its purpose is to give qualitative reasonable conclusions. However, we leave
for future work the model estimation which can be useful to more adequately
examine the impact of financial frictions of forecasting performance.
Estimation also can be useful also in analyzing historical decomposition of
shocks, scenario analyses, and qualitative stress tests of the credit
market. Furthermore, one can expand the empirical analysis using time series
methods (such as VAR, Local Projections) or using loan level data (mortgage
database). Lastly, the model can be analyzed in the presense of the
effective lower bound of the policy interest rate, where in this case we may
get much higher amplification due to the limitation of the monetary policy
to mitigate demand shocks (we analyzed this case but did not put it in
here). This topic is discussed in \cite{Benigno_et_al_2020} and \cite%
{Cohen2022Analyzing} among others.

\phantomsection
\addcontentsline{toc}{section}{\refname}%

\bibliographystyle{elsart-harv}
\bibliography{D:/GitHub/CohenimTools/CohenimTex/BibtexDatabase/bibtex-FinancialfrictionAN.bib}

\begin{thebibliography}{33}
\expandafter\ifx\csname natexlab\endcsname\relax\def\natexlab#1{#1}\fi
\expandafter\ifx\csname url\endcsname\relax
  \def\url#1{\texttt{#1}}\fi
\expandafter\ifx\csname urlprefix\endcsname\relax\def\urlprefix{URL }\fi

\bibitem[{Adrian et~al.(2020)Adrian, Liang, Zabczyk, and
  Duarte}]{Adrian_et_al_2020}
Adrian, T., Liang, N., Zabczyk, P., Duarte, F., 2020. Monetary and
  macroprudential policy with endogenous risk. {{IMF}} Working Papers 2020/236,
  {International Monetary Fund}.

\bibitem[{Argov et~al.(2012)Argov, Barnea, Binyamini, Borenstein, Elkayam, and
  Rozenshtrom}]{argov2012moise}
Argov, E., Barnea, E., Binyamini, A., Borenstein, E., Elkayam, D., Rozenshtrom,
  I., 2012. {{MOISE}}: {{A DSGE Model}} for the {{Israeli Economy}}. Bank of
  Israel Discussion Papers 2012.06.

\bibitem[{Argov and Elkayam(2009)}]{Argov_Elkayam_2009}
Argov, E., Elkayam, D., 2009. An estimated new keynesian model for {{Israel}}.
  Bank of Israel Review (Seker)~82, {Bank of Israel}.

\bibitem[{Benchimol(2016)}]{Benchimol_2016}
Benchimol, J., 2016. Money and monetary policy in {{Israel}} during the last
  decade. Journal of Policy Modeling 38~(1), 103--124.

\bibitem[{Benchimol(2022)}]{Benchimol_2022}
Benchimol, J., 2022. Historical and desirable policy rules in {{Israel}}: {{A
  DSGE}} perspective. Mimeo, {Bank of Israel}.

\bibitem[{Benchimol et~al.(2022)Benchimol, Gamrasni, Kahn, Ribon, Saadon,
  {Ben-Ze'ev}, Segal, and Shizgal}]{Benchimol_et_al_2021}
Benchimol, J., Gamrasni, I., Kahn, M., Ribon, S., Saadon, Y., {Ben-Ze'ev}, N.,
  Segal, A., Shizgal, Y., Jul. 2022. The interaction between domestic monetary
  policy and macroprudential policy in {{Israel}}. Economic Modelling 112,
  105872.

\bibitem[{Benigno et~al.(2020)Benigno, Eggertsson, and
  Romei}]{Benigno_et_al_2020}
Benigno, P., Eggertsson, G.~B., Romei, F., 2020. Dynamic debt deleveraging and
  optimal monetary policy. American Economic Journal: Macroeconomics 12~(2),
  310--350.

\bibitem[{Bernanke et~al.(1999)Bernanke, Gertler, and
  Gilchrist}]{Bernanke_et_al_1999}
Bernanke, B.~S., Gertler, M., Gilchrist, S., 1999. The financial accelerator in
  a quantitative business cycle framework. In: Taylor, J.~B., Woodford, M.
  (Eds.), Handbook of Macroeconomics. Vol.~1 of Handbook of Macroeconomics.
  {Elsevier}, Ch.~21, pp. 1341--1393.

\bibitem[{Boissay et~al.(2022)Boissay, Collard, Gal{\'i}, and
  Manea}]{Boissay.etal2022Monetary}
Boissay, F., Collard, F., Gal{\'i}, J., Manea, C., Jan. 2022. Monetary policy
  and endogenous financial crises. Bank for International Settlements No 991.

\bibitem[{Borio(2014)}]{Borio2014Macroprudential}
Borio, C., 2014. Macroprudential frameworks: ({{Too}}) great expectations?
  VoxEU.org Macroprudentialism, 29.

\bibitem[{Borio et~al.(2018)Borio, Disyatat, Juselius, and
  Rungcharoenkitkul}]{Borio.etal2018Monetarya}
Borio, C., Disyatat, P., Juselius, M., Rungcharoenkitkul, P., 2018. Monetary
  {{Policy}} in the {{Grip}} of a {{Pincer Movement}}. Central Bank of Chile,
  46.

\bibitem[{{Brzoza-Brzezina} et~al.(2013){Brzoza-Brzezina}, Kolasa, and
  Makarski}]{Brzoza_Brzezina_et_al_2013}
{Brzoza-Brzezina}, M., Kolasa, M., Makarski, K., Jan. 2013. The {{Anatomy}} of
  {{Standard Dsge Models}} with {{Financial Frictions}}. Journal of Economic
  Dynamics and Control 37~(1), 32--51.

\bibitem[{{Cesa-Bianchi} and Anderson(2020)}]{Cesa_Bianchi_Anderson_2020}
{Cesa-Bianchi}, A., Anderson, G., 2020. Crossing the credit channel: {{Credit}}
  spreads and firm heterogeneity. {{IMF}} Working Papers 2020/267,
  {International Monetary Fund}.

\bibitem[{Chen~Zion(2021)}]{Chen_Zion_2021}
Chen~Zion, Y., 2021. Estimation of a macroeconomic model for the {{Israeli}}
  economy. Bank of {{Israel}} Working Papers 2021.22, {Bank of Israel}.

\bibitem[{Clarida et~al.(2002)Clarida, Gali, and Gertler}]{CGG_2002}
Clarida, R., Gali, J., Gertler, M., 2002. A simple framework for international
  monetary policy analysis. Journal of Monetary Economics 49~(5), 879--904.

\bibitem[{Cohen(2022)}]{Cohen2022Analyzing}
Cohen, N., Sep. 2022. Analyzing the {{Amplification Mechanisms}} of {{Debt
  Deleveraging}}. Research Department, Bank of Israel 2022~(16).

\bibitem[{C{\'u}rdia and Woodford(2010)}]{Curdia_Woodford_2010}
C{\'u}rdia, V., Woodford, M., Aug. 2010. Credit {{Spreads}} and {{Monetary
  Policy}}. Journal of Money, Credit and Banking 42, 3--35.

\bibitem[{C{\'u}rdia and Woodford(2016)}]{Curdia_Woodford_2016}
C{\'u}rdia, V., Woodford, M., 2016. Credit frictions and optimal monetary
  policy. Journal of Monetary Economics 84~(C), 30--65.

\bibitem[{Gertler and Karadi(2011)}]{Gertler_Karadi_2011}
Gertler, M., Karadi, P., Jan. 2011. A model of unconventional monetary policy.
  Journal of Monetary Economics 58~(1), 17--34.

\bibitem[{Gertler and Karadi(2015)}]{Gertler_Karadi_2015}
Gertler, M., Karadi, P., Jan. 2015. Monetary {{Policy Surprises}}, {{Credit
  Costs}}, and {{Economic Activity}}. American Economic Journal: Macroeconomics
  7~(1), 44--76.

\bibitem[{Gourio et~al.(2018)Gourio, Kashyap, and Sim}]{Gourio.etal2018Trade}
Gourio, F., Kashyap, A.~K., Sim, J.~W., Mar. 2018. The {{Trade}} offs in
  {{Leaning Against}} the {{Wind}}. IMF Economic Review 66~(1), 70--115.

\bibitem[{Iacoviello(2005)}]{Iacoviello_2005}
Iacoviello, M., May 2005. House {{Prices}}, {{Borrowing Constraints}}, and
  {{Monetary Policy}} in the {{Business Cycle}}. American Economic Review
  95~(3), 739--764.

\bibitem[{Ilek and Segal(2022)}]{Ilek_Segal_2022}
Ilek, A., Segal, G., 2022. A simple theory-based estimate of the real natural
  rate of interest in open economies. Bank of {{Israel}} Working Papers
  2022.06, {Bank of Israel}.

\bibitem[{Juselius et~al.(2016)Juselius, Borio, Disyatat, and
  Drehmann}]{Juselius.etal2016Monetary}
Juselius, M., Borio, C.~E., Disyatat, P., Drehmann, M., 2016. Monetary
  {{Policy}}, the {{Financial Cycle}} and {{Ultralow Interest Rates}}. SSRN
  Electronic Journal.

\bibitem[{Kiyotaki and Moore(1997)}]{Kiyotaki_Moore_1997}
Kiyotaki, N., Moore, J., Apr. 1997. Credit {{Cycles}}. Journal of Political
  Economy 105~(2), 211--248.

\bibitem[{Laxton et~al.(2006)Laxton, Berg, and Karam}]{Laxton_et_al2006}
Laxton, D., Berg, A., Karam, P., 2006. A practical model-based approach to
  monetary policy analysis-overview. {{IMF}} Working Papers 06/80,
  {International Monetary Fund}.

\bibitem[{McCulley and Toloui(2008)}]{McCulley_Toloui_2008}
McCulley, P., Toloui, R., 2008. Chasing the neutral rate down: {{Financial}}
  conditions, monetary policy, and the taylor rule. Tech. rep., {PIMCO Global
  Central Bank Focus}.

\bibitem[{{Schmitt-Groh{\'e}} and Uribe(2003)}]{Schmitt_Grohe_and_Uribe_2003}
{Schmitt-Groh{\'e}}, S., Uribe, M., 2003. Closing small open economy models.
  Journal of International Economics 61~(1), 163--185.

\bibitem[{Segal(2007)}]{Segal_2007}
Segal, G., 2007. An optimal discretionary policy rule under rational
  expectations applied to {{Israel}}. Bank of {{Israel}} Working Papers
  2007.05, {Bank of Israel}.

\bibitem[{Shami(2019)}]{Shami_2019}
Shami, L., Dec. 2019. Household {{Debt}} in {{Israel}}. Tech. rep., {Taub
  Center Policy Paper}, {Jerusalem}.

\bibitem[{Svensson(2017)}]{Svensson2017Costbenefit}
Svensson, L.~E., Oct. 2017. Cost-benefit analysis of leaning against the wind.
  Journal of Monetary Economics 90, 193--213.

\bibitem[{Svensson(2018)}]{Svensson2018Monetary}
Svensson, L.~E., 2018. Monetary policy and macroprudential policy:
  {{Different}} and separate? The Canadian Journal of Economics / Revue
  canadienne d'Economique 51~(3), 802--827.

\bibitem[{Svensson(2014)}]{Svensson2014Inflation}
Svensson, L. E.~O., 2014. Inflation {{Targeting}} and "{{Leaning}} against the
  {{Wind}}". International Journal of Central Banking, 12.

\end{thebibliography}
\appendix
\setcounter{equation}{0}
\renewcommand{\theequation}{\Alph{section}.\arabic{equation}}
\setcounter{theorem}{0}
\renewcommand{\thetheorem}{\Alph{section}.\arabic{theorem}}%

\part{Appendix}

\section{Empirical evidence of financial friction in the mortgage market in
Israel\label{app: empirical evidence}}

As discussed in \ref{subsubsec: spread elasticity},\ here we will validate
empirically our assumption in the model that there is a financial friction
in the mortgage market in Israel, which is reflected in a positive
relationship between the interest rate spread and the leverage ratio. We
apply three tests to validate the positive relationship mentioned above.

\subsection{\protect\Large Test 1}

In the first version of Eq. \ref{reg spread lev} explained in \ref%
{subsubsec: spread elasticity}\ we use a spread, $spread_{t}^{H}=r^{m%
\_w}-r^{10\_bond}$, where $r^{m\_w}$ is a weighted average (real) interest
rate on mortgages\footnote{%
It is calculated by the Information \ and Statistics Department at the BOI.
The calculation takes into account both interest rate on indexed and
non-indexed mortgages and different maturities.} and $r^{10\_bond}$ is a
real yield on government bonds for 10 years to maturity. In the second
version we use a spread, $spread_{t}^{H}=r^{m}-r^{bank},$ where $r^{m}$ is a
fixed real interest rate on mortgages and $r^{bank}$ is a real yield on
bonds of commercial banks. In the last version we use a spread,$%
spread_{t}^{H}=i^{m}-i^{bank},$ where $i^{m}$ is a fixed nominal interest
rate on mortgages and $i^{bank}$ is a nominal yield on bonds of commercial
banks. The estimation results are shown below:

\textbf{Version 1:\footnote{%
The reported p-values in all equations of Appendix A are based on the S.E.
of the parameters corrected by the Newey-West methodology.}}%
\begin{eqnarray}
spread_{t}^{H} &=&\underset{(0.00)}{-5.89}+\underset{(0.00)}{0.07}%
Lev_{t}^{H}-\underset{(0.00)}{0.16}\pi _{t}^{H}-\underset{(0.15)}{0.06}\pi
_{t-1}^{H}-\underset{(0.06)}{0.07}\pi _{t-2}^{H}-\underset{(0.10)}{0.06}\pi
_{t-3}^{H},\text{ }  \notag \\
R^{2} &=&0.74,\text{ }DW=0.59  \label{ver1}
\end{eqnarray}

\textbf{Version 2:}%
\begin{eqnarray}
spread_{t}^{H} &=&\underset{(0.00)}{-11.15}+\underset{(0.00)}{0.13}%
Lev_{t}^{H}+\underset{(0.92)}{0.005}\pi _{t}^{H}-\underset{(0.13)}{0.08}\pi
_{t-1}^{H}-\underset{(0.23)}{0.05}\pi _{t-2}^{H}\underset{(0.61)}{+0.03}\pi
_{t-3}^{H},\text{ }  \notag \\
R^{2} &=&0.66,\text{ }DW=0.36  \label{ver2}
\end{eqnarray}

\textbf{Version 3:}%
\begin{eqnarray}
spread_{t}^{H} &=&\underset{(0.06)}{-3.13}+\underset{(0.00)}{0.05}%
Lev_{t}^{H}-\underset{(0.15)}{0.10}\pi _{t}^{H}-\underset{(0.66)}{0.02}\pi
_{t-1}^{H}-\underset{(0.08)}{0.10}\pi _{t-2}^{H}\underset{(0.39)}{+0.09}\pi
_{t-3}^{H},\text{ }  \notag \\
R^{2} &=&0.31,\text{ }DW=0.30  \label{ver3}
\end{eqnarray}

We can see\ from Eq's \ref{ver1}--\ref{ver3} that the effect of the leverage
ratio on various spreads lies between $0.05$ to $0.13$ and is always highly
significant. As expected, the collateral effect of home prices, $\pi
_{t}^{H}...\pi _{t-3}^{H}$, on the spread is negative (and significant) in
most cases.

As we mentioned previously, the estimated parameter $\beta $ can be upward
biased because macroprudential steps applied during the sample period could
increase the elasticity of spread to leverage. To see how the basic
parameter $\beta $ changes, we add to the regression additional explanatory
variable $\gamma Lev_{t}^{H}\cdot dum_{t},$ where $dum_{t}$ is a dummy
variable capturing the macroprudential measures. Under this specification
the elasticity of the spread to the leverage is $\beta +\gamma \cdot
dum_{t}. $ We consider three types of variable $dum_{t}$ (all of them are
taken from \cite{Benchimol_et_al_2021}). The first type of the dummy
variables is in a form of 'steps', so when a new macroprudential measure was
implemented in the credit housing market, the new 'step' in the dummy
variable is added. The second type of the dummy variable is based on the
same macroprudential measures as before, but now the dummy receives $1$ when
a new measure was implemented in some quarter and $0$ otherwise. This dummy
is less reasonable and barely significant in estimation, because it assumes
only very transitory effect on the elasticity and the spread. The third type
of the dummy variable is based on general macroprudential measurers
constructed by \cite{Benchimol_et_al_2021}. We obtain that for all three
versions of the dummy variable the parameter $\gamma $ is positive but it is
significant only for the first dummy, which to some extent validates our
concern of upward bias of $\beta $. Most importantly, we find no evidence
that inclusion of $\gamma Lev_{t}^{H}\cdot dum_{t}$ wipes out the parameter $%
\beta .$ It is still positive and indeed smaller in some cases than in the
original estimation.

\subsection{\protect\Large Test 2}

Now we use an alternative measure of the leverage ratio $\overset{--}{%
Lev_{t}^{H}}=\frac{B_{t}^{H\_new}}{Y_{t}},$ where $B_{t}^{H\_new}$ is a
volume of new mortgages in quarter $t$ (instead of stock of mortgages) and $%
Y_{t}$ is a GDP (with seasonal adjustment) in quarter $t$ (sample
2004Q1-2021Q3):%
\begin{equation}
spread_{t}^{H}=c+\beta \overset{--}{Lev_{t}^{H}}+\alpha _{0}\pi
_{t}^{H}+\alpha _{1}\pi _{t-1}^{H}+\alpha _{2}\pi _{t-2}^{H}+\alpha _{3}\pi
_{t-3}^{H}  \label{test2}
\end{equation}

The estimation results of three versions of spread measure are shown below.

\textbf{Version 1:}%
\begin{eqnarray}
spread_{t}^{H} &=&\underset{(0.03)}{-0.84}+\underset{(0.00)}{0.47}\overset{--%
}{Lev_{t}^{H}}-\underset{(0.00)}{0.32}\pi _{t}^{H}-\underset{(0.00)}{0.11}%
\pi _{t-1}^{H}-\underset{(0.07)}{0.08}\pi _{t-2}^{H}-\underset{(0.01)}{0.10}%
\pi _{t-3}^{H},  \notag \\
R^{2} &=&0.66,\text{ }DW=0.78  \label{A6}
\end{eqnarray}

\textbf{Version 2:}%
\begin{eqnarray}
spread_{t}^{H} &=&\underset{(0.00)}{-1.76}+\underset{(0.00)}{0.88}\overset{--%
}{Lev_{t}^{H}}-\underset{(0.00)}{0.30}\pi _{t}^{H}-\underset{(0.00)}{0.19}%
\pi _{t-1}^{H}-\underset{(0.29)}{0.07}\pi _{t-2}^{H}-\underset{(0.49)}{0.05}%
\pi _{t-3}^{H},  \notag \\
R^{2} &=&0.52,\text{ }DW=0.46  \label{A7}
\end{eqnarray}

\textbf{Version 3:}%
\begin{eqnarray}
spread_{t}^{H} &=&\underset{(0.67)}{0.26}+\underset{(0.00)}{0.39}\overset{--}%
{Lev_{t}^{H}}-\underset{(0.00)}{0.23}\pi _{t}^{H}-\underset{(0.17)}{0.07}\pi
_{t-1}^{H}-\underset{(0.08)}{0.11}\pi _{t-2}^{H}+\underset{(0.52)}{0.06}\pi
_{t-3}^{H},  \notag \\
R^{2} &=&0.32,\text{ }DW=0.41  \label{A8}
\end{eqnarray}

We see from Eq's (\ref{A6})-(\ref{A8}) that the effect on new loans to GDP
is much higher than in Test 1 as expected. The collateral effect of home
prices on spreads is still at place.

\subsection{\protect\Large Test 3}

We also tested the relationship between the mortgage interest rate and LTV
ratio, where $LTV_{t}=\frac{B_{t}^{H\_new}}{P_{t}^{H}}$ ($B_{t}^{H\_new}$ is
a volume of new mortgages, $P_{t}^{H}$ is a home price). We compared the
average nominal (and real) interest rate of loans with LTV of 30-45\%,
45-60\% and 60-75\% in sample 2011M7-2021M11. We found that the average
nominal (and real) interest rate of loans with LTV of 45-60\% is higher by
0.3 p.p. compared to the loans with LTV of 45-60\% and a similar gap exists
between the LTV of 60-75\% and LTV of 45-60\%. That means, broadly speaking,
that increase in LTV by 15\% increases the interest rate by 0.3 p.p.,
implying elasticity of the mortgage interest rate with respect to LTV of
about 0.02. We notice, however, that higher LTV is associated with longer
maturity of loans in the data, therefore it could be that the higher
maturity induces higher interest rates and not necessarily higher LTV. We
assert, however, that higher maturity is also a part of the risk for the
banks (lenders) because higher maturity increases a probability of default
of the borrowers during the maturity period.\footnote{%
Clearly, say, an NIS 1 million loan for 10 years is riskier than \ the same
loan for 1 year, given the same leverage ratio of the borrower.} In summary,
it is unimportant for our purposes to identify whether higher interest rate
is due to higher LTV or due to higher maturity. Both factors reflect risk of
the borrowers and therefore induce a positive relationship between interest
rate and risk factors, which is a heart of financial friction we consider in
our paper.

\section{Deriving elasticity for our model\label{app: deriving elasticity}}

The last step is to translate the estimated elasticity in the mortgage
market to elasticity in terms of total debt of the households which contains
both mortgages and non-mortgage credit. For that we need to examine
spread-leverage ratio in non-mortgage credit market as well. Unfortunately
no reliable examination is feasible since the data is very limited. We adopt
a conservative approach and assume that the spread-leverage relationship in
non-mortgage credit market is close to null. For the housing market we use
elasticity from Eq. \ref{ver1} which is most suitable to our model.

Notice that by definition the total debt is a sum of a mortgage and
non-mortgage debt%
\begin{equation}
B_{t}=B_{t}^{H}+B_{t}^{N\_H}  \label{A9}
\end{equation}

The interest rate payment on the total debt is a sum of interest rate
payments of mortgage and non-mortgage debt%
\begin{equation}
(R_{t}+spread_{t})B_{t}=(R_{t}+spread_{t}^{H})B_{t}^{H}+(R_{t}+spread_{t}^{N%
\_H})B_{t}^{N\_H}  \label{A10}
\end{equation}

where $R_{t}$ is a riskless interest rate, $spread_{t}$ is a spread on total
credit, $spread_{t}^{H}$ is a spread on mortgage credit and $%
spread_{t}^{N\_H}$ is a spread on non-mortgage credit. Exploiting Eq's \ref%
{A9} and \ref{A10} we get%
\begin{equation*}
spread_{t}=spread_{t}^{H}\frac{B_{t}^{H}}{B_{t}}+spread_{t}^{N\_H}\frac{%
B_{t}^{N\_H}}{B_{t}}
\end{equation*}

Using the fact that in Israel a weight of mortgages in total credit of
households is 2/3, and assuming elasticity from Eq. \ref{ver1} we rewrite
the previous equation 
\begin{equation}
spread_{t}=(spread^{H}+\underset{0.07}{\underbrace{\beta ^{H}}}\overset{%
\symbol{126}}{Lev_{t}^{H}})\underset{2/3}{\underbrace{\frac{B_{t}^{H}}{B_{t}}%
}}+(spread^{N\_H}+\underset{\symbol{126}0}{\underbrace{\beta ^{N\_H}}}%
\overset{\symbol{126}}{Lev_{t}^{N\_H}})\underset{1/3}{\underbrace{\frac{%
B_{t}^{N\_H}}{B_{t}}}}  \label{A11}
\end{equation}

where $spread^{H}$ and $spread^{H}$ are SS values of spread on mortgage
credit and non-mortgage credit, respectively. $\overset{\symbol{126}}{%
Lev_{t}^{H}}$ is a deviation of spread on mortgage credit from its SS value.
After some simplification on Eq. \ref{A11} we get%
\begin{equation}
spread_{t}=\frac{2}{3}(spread^{H}+0.07\overset{\symbol{126}}{Lev_{t}^{H}})+%
\frac{1}{3}(spread^{N\_H})  \label{A12}
\end{equation}%
Note that 
\begin{equation*}
Lev_{t}^{H}=\frac{B_{t}^{H}}{Y_{t}}\frac{B_{t}}{B_{t}}=\frac{2}{3}Lev_{t}
\end{equation*}%
where $Lev_{t}=\frac{B_{t}}{Y_{t}}.$ Therefore the following condition also
holds (since in SS, $Lev^{H}=$ $\frac{2}{3}Lev$)%
\begin{equation}
\overset{\symbol{126}}{Lev_{t}^{H}}=\frac{2}{3}\overset{\symbol{126}}{Lev_{t}%
}  \label{A13}
\end{equation}

Substituting Eq. \ref{A13} into Eq. \ref{A12} we get%
\begin{equation*}
spread_{t}=\frac{2}{3}(spread^{H}+0.07(\frac{2}{3})\overset{\symbol{126}}{%
Lev_{t}})+\frac{1}{3}(spread^{N\_H})
\end{equation*}

From Eq. \ref{A11} we see that in SS, $spread=\frac{2}{3}(spread^{H})+\frac{1%
}{3}(spread^{N\_H}),$ therefore \ 
\begin{equation*}
spread_{t}=spread+0.07(\frac{2}{3})^{2}\overset{\symbol{126}}{Lev_{t}}
\end{equation*}

Finally, we obtain the relationship between the interest rate spread gap and
leverage gap as appears in the model. 
\begin{equation}
\overset{\symbol{126}}{spread_{t}}=0.031\overset{\symbol{126}}{Lev_{t}}
\label{A14}
\end{equation}

\section{Foreign shock IRF\label{app: Foreign shock IRF}}

\begin{figure}
\caption{\textbf{IRF to positive shock to foreign monetary policy of 1 SE}}
\includegraphics[width=\linewidth]{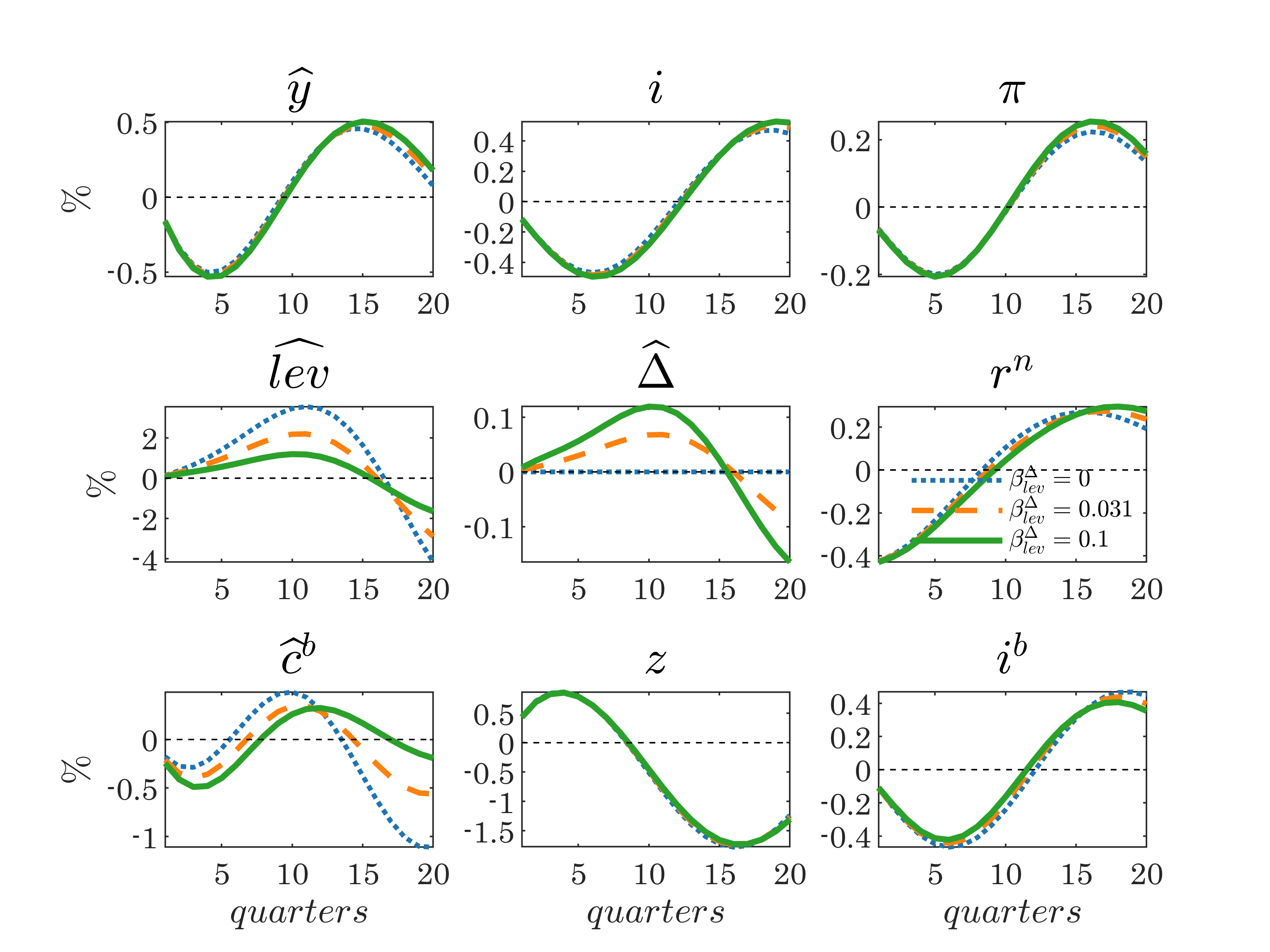}
\label{fig: IRF monetary policy row}
{\footnotesize
\textit{Notes:} Variables as deviation from steady state: output gap, inflation, interest rate,
leverage, credit spread, borowers consumption, real exchange rate and borowers' interest rate.
Blue dotted line: represents the case without financial friction. 
Orange line: represents the baseline case with financial friction, but without macroprudential policy 
(Elasticity  of spread to leverage is 0.031).
Green line: represents the case with macroprudential policy.
(elasticity  of spread to leverage is 0.1).
All simulations are under specification of low borowers' aversion to leverage, v=0.0225
}
\end{figure}

Foreign monetary shock see Fig. 
\ref{fig: IRF monetary policy row}%
.

\section{IRF under Financial Accelerator Calibration \label{app: IRF
Financial Accelerator}}

\begin{figure}
\caption{\textbf{IRF to positive shock to monetary policy of 1 SE}}
\includegraphics[width=\linewidth]{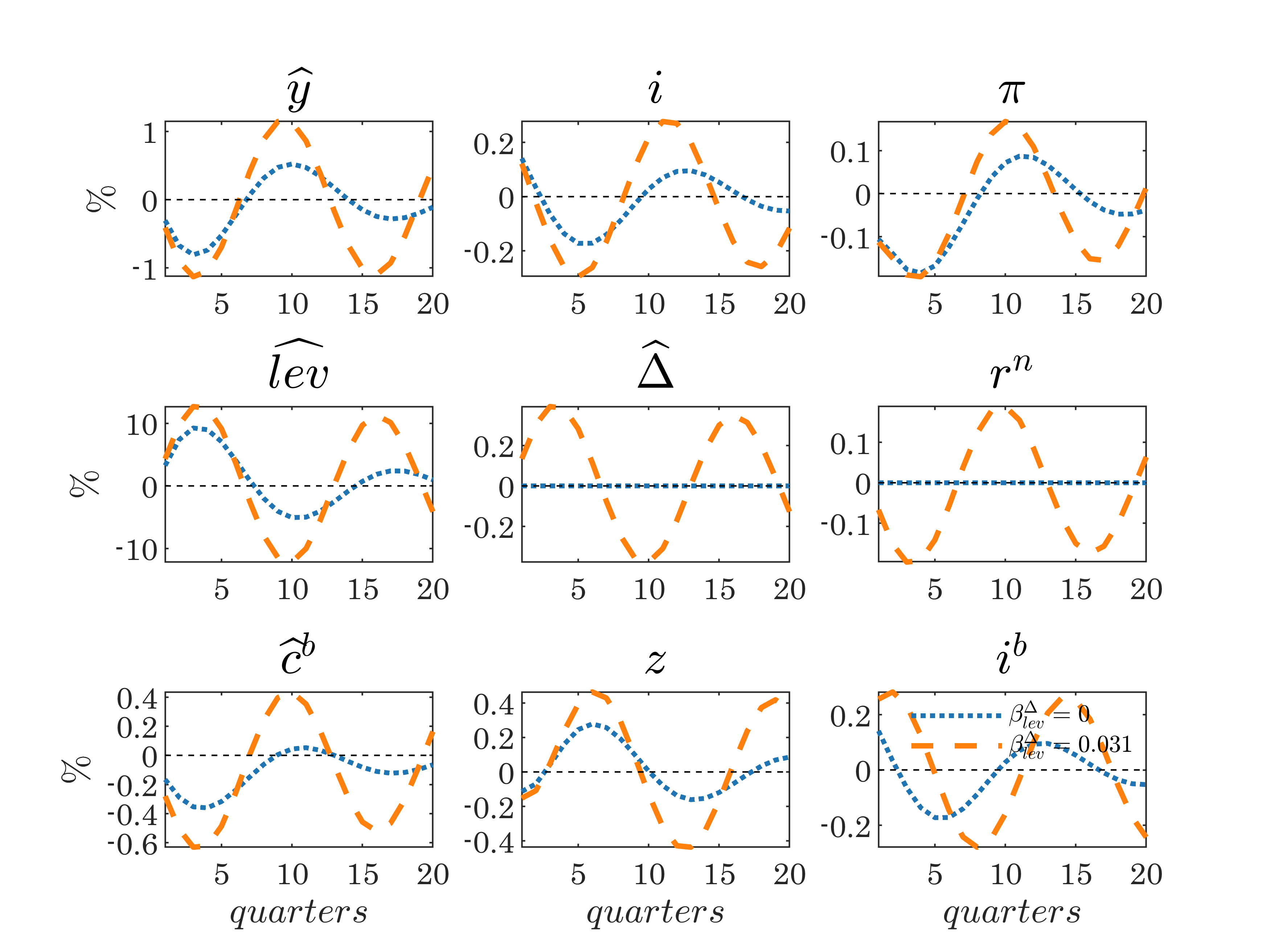}
\label{fig: IRF monetary policy FA}
{\footnotesize
\textit{Notes:} Variables as deviation from steady state: output gap, inflation, interest rate,
leverage, credit spread, borowers consumption, real exchange rate and borowers' interest rate.
Blue dotted line: represents the case without financial friction. 
Orange line: represents the baseline case with financial friction, but without macroprudential policy 
(Elasticity  of spread to leverage is 0.031).
All simulations are under specification of low borowers' aversion to leverage, v=0.0225
}
\end{figure}

As can be seen in Fig. 
\ref{fig: IRF monetary policy FA}
it is possible to have financial accelerator dynamic under some model
calibration. This includes calibration of the\ elasticity of the output gap
to NRI: $\beta _{r}^{y}=-0.5$ (see Eq. \ref{ygap}), which is in the upper
range of several estimates of this parameter exists for Israel. But we also
need to calibrate $\beta _{r}^{c^{b}}=0.2$ (instead of $5$) which is much
smaller from any value acceptable for Israel economy (economically meaning
that the consumption of the borrowers is less sensitive to the interest rate
than the lenders, in opposite to \cite{Curdia_Woodford_2016}). In summery,
the model may give a financial acelerator dynamic, but not for a reasonable
parameters for Israel.

\end{document}